\documentclass{article}

\usepackage{arxiv}

\usepackage[utf8]{inputenc} 
\usepackage[T1]{fontenc}    
\usepackage{hyperref}       
\usepackage{url}            
\usepackage{booktabs}       
\usepackage{amsfonts}       
\usepackage{nicefrac}       
\usepackage{microtype}      
\usepackage{lipsum}

\usepackage{amsfonts}
\usepackage{amsmath}
\usepackage{subcaption}
\usepackage{graphicx}
\usepackage{amssymb}
\usepackage[inkscapelatex=false]{svg}
\usepackage{mathtools}
\usepackage{xcolor}
\usepackage{newtxtext,newtxmath}
\usepackage{lipsum}
\usepackage{float}
\usepackage{algorithm}
\usepackage{algpseudocode}
\usepackage{stmaryrd}
\usepackage{algorithm}
\usepackage{algpseudocode}

\newcommand{\R}{{\mathbb{R}}}

\newcommand{\N}{{\mathbb{N}}}
\newcommand{\No}{{\mathcal{N}}}
\newcommand{\Lo}{{\mathbb{L}}}

\newcommand{\X}{{\mathbf{X}}}

\newcommand{\W}{{\mathbf{W}}}
\newcommand{\w}{{\mathsf{w}}}
\newcommand{\Wo}{{\mathcal{W}}}
\newcommand{\U}{{\mathbf{U}}}

\newtheorem{theorem}{Theorem}[section]
\newtheorem{assumption}{Assumption}

\newtheorem{definition}[theorem]{Definition}
\newtheorem{lemma}[theorem]{Lemma}
\newtheorem{remark}[theorem]{Remark}
\newenvironment{proof}{\paragraph{Proof:}}{\hfill$\square$}
\newtheorem{problem}[theorem]{Problem}
\usepackage{graphicx} 

\title{Neural Controller for Incremental Stability of Unknown Continuous-time Systems}

\author{
 Ahan Basu \\
  Centre for Cyber-Physical Systems\\
  IISc, Bengaluru, India\\
  \texttt{ahanbasu@iisc.ac.in} \\
   \And
 Bhabani Shankar Dey \\
  Centre for Cyber-Physical Systems\\
  IISc, Bengaluru, India\\
  \texttt{bhabanishan1@iisc.ac.in} \\
  \And
 Pushpak Jagtap \\
  Centre for Cyber-Physical Systems\\
  IISc, Bengaluru, India\\
  \texttt{pushpak@iisc.ac.in} \\
}

\begin{document}

\maketitle

\begin{abstract}
This work primarily focuses on synthesizing a controller that guarantees an unknown continuous-time system to be incrementally input-to-state stable ($\delta$-ISS). In this context, the notion of $\delta$-ISS control Lyapunov function ($\delta$-ISS-CLF) for the continuous-time system is introduced. Combined with the controller, the $\delta$-ISS-CLF guarantees that the system is incrementally stable. As the paper deals with unknown dynamical systems, the controller as well as the $\delta$-ISS-CLF are parametrized using neural networks. The data set used to train the neural networks is generated from the state space of the system by proper sampling. Now, to give a formal guarantee that the controller makes the system incrementally stable, we develop a validity condition by having some Lipschitz continuity assumptions and incorporate the condition into the training framework to ensure a provable correctness guarantee at the end of the training process. Finally, we demonstrate the effectiveness of the proposed approach through several case studies: a scalar system with a non-affine, non-polynomial structure, a one-link manipulator system, a nonlinear Moore-Greitzer model of a jet engine, a magnetic levitator system and a rotating rigid spacecraft model.
\end{abstract}

\section{Introduction}\label{sec:intro}
Input-to-state stability has been a promising tool for analyzing and studying robust stability questions for nonlinear control systems. This traditional stability analysis mainly focuses on the study of the system's trajectory to converge into an equilibrium point or some other nominal trajectory. Compared to these studies, incremental stability proposes a stronger notion of stability, denoting convergence of arbitrary trajectories towards each other rather than to a particular point \cite{angeli2002lyapunov}. As a result, it has gained significant attention for its broad applicability, including nonlinear analog circuit modeling \cite{analog}, cyclic feedback system synchronization \cite{cyclic_feedback}, symbolic model development \cite{symbolic1,zamani2017towards,jagtap2020symbolic,jagtap2017quest,sadek2023compositional}, stability of interconnected systems \cite{DD-Stability}, oscillator synchronization \cite{inter_osci}, and complex network analysis \cite{synchComplex}.

Several promising tools have been developed in the last decades to analyze the property of incremental stability. Contraction analysis \cite{Contraction, sun2021learning, tsukamoto} and convergent dynamics \cite{Conv_dyn} have emerged as promising tools to analyze incremental stability properties for nonlinear systems. However, traditional stability analysis is done using comparison functions, which leads to the notion of invariance of the system with respect to change in coordinates \cite{grune1999asymptotic}. This leads the research community to find the most powerful tool to analyze incremental stability, namely incremental Lyapunov functions \cite{angeli2002lyapunov,DT-ISS_prop,zamani2011lyapunov}. Later, these tools have also been extended to analyze the incremental stability of a wide class of systems, such as nonlinear systems \cite{angeli2002lyapunov}, stochastic systems \cite{biemond2018incremental}, hybrid dynamical systems \cite{biemond2018incremental}, time-delayed systems \cite{chaillet2013razumikhin}, and interconnected switched systems \cite{dey2023incremental}. 

Incremental stability is a powerful tool for the construction of a finite abstraction of nonlinear control systems \cite{girard2009approximately}, which motivates the design of controllers that enforce it. However, most of the existing approaches to design controllers ensuring incremental stability \cite{zamani2011backstepping, zamani2013backstepping, jagtap2017backstepping} rely on the complete knowledge of the dynamics of the systems, making them unsuitable when uncertainties or modelling errors are present in real-world scenarios. {To address these limitations posed by unknown dynamics or modelling errors, data-driven and learning-based approaches have gained significant attention. The authors in \cite{sangeerth2025controller,sundarsingh2024backstepping} employ the Gaussian process to collect data from the system and develop a backstepping-like controller. However, this method assumes the system to be control-affine and then designs the controller.  Compared to it, a very recent approach for designing a controller to achieve incremental ISS \cite{zaker2024controller} utilizes a data-driven technique, but is only limited to polynomial-type structured dynamics, where data is extracted from system trajectories. However, both these works need the system to be identified first before the control design step, which is indeed challenging.}

Among the learning-based approaches for verification and controller synthesis, deep learning-based techniques have gained prominence to estimate system dynamics or synthesize controllers for different task specifications alongside constructing Lyapunov or barrier functions to guarantee stability or safety. The existing literature \cite{formal_nn_Lyapunov,Neural_CBF,tayal2024learning} addresses the issue of safety and stability while eliminating the need for explicit knowledge of the system dynamics. Leveraging the universal approximation capabilities of neural networks, functions can be directly synthesized. {Some recent works have used neural networks adaptively to learn the controller in an iterative fashion \cite{shen2024neural} while data-driven control via Q-learning \cite{shen2024event} has been employed to solve tracking problems as well.} However, the key challenge to use neural networks as Lyapunov functions or controllers is to provide a formal guarantee, as the training relies on discrete samples, which cover only a limited part of the continuous state space. {While falsification-based approaches \cite{chang2019neural} have been proposed to mitigate this by adaptively sampling ``difficult” regions, they provide only probabilistic confidence; in contrast, we try to enforce deterministic satisfaction of $\delta$-ISS conditions through constrained optimization. The CEGIS-based approach \cite{abate2020formal} has been a powerful tool as well to verify neural Lyapunov function, but it works in a post-hoc verification scenario, while in this work, we incorporate verification within the training procedure without the need for any other solver.}

\textit{Contributions}: Building on our earlier work on formally verified neural network controllers for unknown discrete-time systems \cite{basu2025controller}, we extend the framework to unknown continuous-time systems. {We propose the notion of an incremental input-to-state stable control Lyapunov function ($\delta$-ISS-CLF) for continuous-time systems and prove that the existence of such a function under the controller leads the closed-loop system to be incrementally stable.} For a general unknown system, we propose data-driven techniques to synthesize $\delta$-ISS-CLF and the controller jointly and hence the notion of $\delta$-ISS-CLF over compact sets is proposed while ensuring forward invariance by leveraging the notion of CBF as seen in \cite{basu2025controller}. We unify the $\delta$-ISS-CLF and CBF conditions in a robust optimization program (ROP) and propose the controller and the $\delta$-ISS-CLF as neural networks, as system dynamics is unknown and only a black-box model is provided, and formulate a scenario convex problem corresponding to the ROP using some Lipschitz continuity assumptions. We propose a validity condition that gives a formal guarantee that the neural controller can make the system incrementally stable and incorporate it in the training framework to avoid any posteriori verification. We validate the effectiveness of our approach by applying it to multiple case studies.

The major contribution of this article over the discrete-time version is twofold. First, the paper highlights the notion of $\delta$-ISS-CLF for continuous time systems over compact sets, which is completely new compared to the existing literature. Second, the derivative of $\delta$-ISS-CLF to ensure $\delta$-ISS needs to be bounded to provide a deterministic guarantee over the controller, for which it requires the derivative of a neural network to be bounded by some Lipschitz constant. In this work, we propose a novel loss function that ensures the derivative of a multi-layer feedforward neural network is also Lipschitz continuous.

The organization of the paper is as follows. In Section \ref{sec:Prob}, we recall the notion of incremental input-to-state stability and introduce the notion of incremental input-to-state stable control Lyapunov function for compact sets. In Section \ref{sec:Neural Lyapunov}, we propose a learning framework that jointly synthesizes a controller and a verifiably correct incremental input-to-state stable control Lyapunov function, both parametrized as neural networks, posing the problem of synthesizing the controller as an optimization problem and solve it with the help of neural network training, as discussed in Section \ref{sec:Training}. Finally, in Section \ref{sec:Cases}, we validate the proposed approach by showcasing multiple case studies and providing a general discussion on the usefulness of the results.

\section{Preliminaries and Problem Formulation}\label{sec:Prob}

\subsection{Notations}
The symbols $\N$, $ \N_0$, $ \R$, $\R^+$, and $\R_0^+$ denote the set of natural, nonnegative integers, real, positive real, and nonnegative real numbers, respectively. 
A vector space of real matrices with $ n $ rows and $ m $ columns is denoted by $\R^{n\times m} $. The column vector space with $n$ rows is represented by $ \R^{n}$.
The Euclidean norm is represented using $|\cdot |$. 
Given a function $\varphi: \R_0^+ \rightarrow \R^m$, its sup-norm (possibly $\infty$-norm) is given by $\lVert \varphi \rVert = \sup\{|\varphi(k)| : k \in \R_0^+\}$.
For $a, b \in \N_0$ with $a \leq b$, the closed interval in $\N_0$ is denoted as $[a; b]$.  
A vector $x \in \mathbb{R}^{n}$ with entries $x_1, \ldots, x_n$ is represented as $[x_1, \ldots, x_n]^\top$, where $x_i \in \mathbb{R}$ denotes the $i$-th element of the vector and $i \in [1;n]$.
The space of diagonal matrices in $\R^{n\times n}$ with nonnegative entries is denoted by $\mathcal{D}_{\geq 0}^n$.
Given a matrix $M\in\R^{n\times m}$, $M^\top$ represents the transpose of matrix $M$. 
A continuous function $\alpha: \R_0^+ \rightarrow \R_0^+$ is said to be class $\mathcal{K}$, if $\alpha(s)>0 $ for all $s>0$, strictly increasing and $\alpha(0)=0$. It is class $\mathcal{K}_\infty$ if it is class $\mathcal{K}$ and $\alpha(s)\rightarrow\infty$ as $s\rightarrow\infty$.
A continuous function $\beta: \R_0^+ \times \R_0^+ \rightarrow \R_0^+$ is said to be a class $\mathcal{KL}$ if $\beta(s,t)$ is a class $\mathcal{K}$ function with respect to $s$ for all $t$ and for fixed $s>0$, $\beta(s,t) \rightarrow 0 $ if $t\rightarrow \infty$. 
For any compact set $\mathcal{C}, \partial \mathcal{C}$ and $int(\mathcal{C})$ denote the boundary and interior of the set $\mathcal{C}$, respectively.
The gradient of a function $f:\X \rightarrow \R$ is denoted by $\nabla f(x)$.
A function $f:\X \rightarrow \R$ is said to $\vartheta$-smooth if for all $x,y \in \X, |\nabla f(x) - \nabla f(y)| \leq \vartheta|x-y|$.
The partial differentiation of a function $f: \X \times \hat{\X} \rightarrow \R$ with respect to a variable $x \in \X$ is given by $\frac{\partial f}{\partial x}$. 

\subsection{Incremental Stability for Continuous-time systems}
We consider a non-linear continuous-time control system (ct-CS) represented by the tuple $\Xi = (\X, \U, f)$ and describe it using the differential equation as
\begin{equation}\label{eq:act_system}
    \dot{\mathsf{x}} = f(\mathsf{x}, \mathsf{u}),
\end{equation}
where $\mathsf{x}(t) \in \X \subseteq \R^n$ is the state of the system and $\mathsf{u}(t) \in \U \subseteq \R^m$ is the input to the system at time $t\geq 0$, the function $f: \R^n \times \R^m \rightarrow \R^n$ is assumed to be locally Lipschitz continuous to ensure the existence and uniqueness of the solution \cite{khalil2002nonlinear}.

Next, the notion of the closed-loop continuous-time control system under a Lipschitz continuous feedback controller $g$ is defined, which is represented as $\Xi_g = (\X, \W, \U, f, g)$, where $\X \subseteq \R^n$ is the state-space of the system, $\U \subseteq \R^m$ is the internal input set of the system, $\W \subseteq \R^p$ is the external input set of the system, $g:\X \times \W \rightarrow \U$ and $f:\X \times \U \rightarrow \R^n$  are maps describing the state evolution as:
\begin{equation}\label{eq:system}
     \dot{\mathsf{x}} = f(\mathsf{x}, g(\mathsf{x}, \mathsf{w})),
\end{equation}
where $\mathsf{x}(t) \in \X$ and $\mathsf{w}(t) \in \W$ are the state and external input of the closed-loop system at time instance $t$, respectively.

Let $\mathsf{x}_{x,\mathsf{w}}(t)$ be the state of the closed-loop system \eqref{eq:system} at time $t$ starting from the initial condition $x\in \X$ under the input signal $\mathsf{w}$. Next, we define the notion of incremental input-to-state stability for the closed-loop continuous-time system \eqref{eq:system}.
\begin{definition}[$\delta$-ISS \cite{angeli2002lyapunov}] \label{def:inc-stable_iss}
    The closed-loop ct-CS in \eqref{eq:system} is incrementally input-to-state stable ($\delta$-ISS) if there exists a class $\mathcal{KL}$ function $\beta$ and a class $\mathcal{K}_\infty$ function $\gamma$, such that for any $t \geq 0$, for all $x, \hat x \in \X $ and any external input signal $\mathsf{w},\hat{\mathsf{w}}$ the following holds:
    \begin{equation}\label{eq:gas-system}
        |\mathsf{x}_{x,\mathsf{w}}(t)-\mathsf{x}_{\hat x,\hat{\mathsf{w}}}(t)| \leq \beta(|x-\hat x|,t) + \gamma(\lVert \mathsf{w} - \hat{\mathsf{w}} \rVert).
    \end{equation}
\end{definition}
If $\mathsf{w} = \hat{\mathsf{w}}=0$, one can recover the notion of incremental global asymptotic stability as defined in \cite{angeli2002lyapunov}. 
Next, we introduce the concept of incremental input-to-state stable control Lyapunov function ($\delta$-ISS-CLF) for continuous-time systems.
\begin{definition}[$\delta$-ISS-CLF]\label{def:ISS-Lf_gen}
    A smooth function $V:\R^n \times \R^n \rightarrow \R_0^+$ is said to be a $\delta$-ISS control Lyapunov function ($\delta$-ISS-CLF) for closed-loop system $\Xi_g = (\R^n, \R^p, \R^m, f, g)$ with the feedback controller $g:\R^n \times \R^p \rightarrow \R^m$, if there exist class $\mathcal{K}_\infty$ functions $\alpha_1, \alpha_2, \sigma$ and a constant $\kappa \in \R^+$ such that:
    \begin{enumerate}\label{cond:ISS-Lf_gen}
        \item[(i)] for all  $x,\hat{x}\in \R^n$: $\alpha_1(|x-\hat{x}|) \leq V(x,\hat{x}) \leq \alpha_2(|x-\hat{x}|),$
        \item[(ii)] for all $x,\hat{x}\in \R^n$ and for all $w, \hat{w} \in \R^p$: $\frac{\partial V}{\partial x}f(x, g(x,w)) + \frac{\partial V}{\partial \hat{x}}f(\hat{x}, g(\hat{x},\hat{w})) \leq -\kappa V(x, \hat{x}) + \sigma(|w - \hat{w}|)$.
    \end{enumerate}
\end{definition}  
Intuitively, the $\delta$-ISS-CLF quantifies the contraction behavior between any two system trajectories. The first condition enforces positive semidefiniteness with respect to the state difference, while the second condition guarantees exponential decay of this difference along the closed-loop trajectories. It should be carefully noted that the proposed $\delta$-ISS-CLF notion differs from classical incremental Lyapunov functions by explicitly incorporating a feedback control law. This allows the function to serve as a constructive tool for controller synthesis. Now, the following theorem describes $\delta$-ISS of the closed-loop system in terms of the existence of a $\delta$-ISS-CLF.
\begin{theorem}\label{th:admit_overall}
    The closed-loop continuous-time control system \eqref{eq:system} will be $\delta$-ISS with respect to external input $\w$ if there exists a $\delta$-ISS-CLF satisfying the conditions of Definition \ref{def:ISS-Lf_gen}.
\end{theorem}
\begin{proof}
     The proof follows similar to that of Theorem 2 of \cite{angeli2002lyapunov}.
\end{proof}

In this work, we aim to solve the problem of incremental stability using a data-driven approach. This requires the collection of data from compact sets; hence, we need the notion of $\delta$-ISS-CLF for compact sets. To do so, we first recall the notion of control forward invariance.
\begin{definition}[Control Forward Invariant Set \cite{liu2019compositional}]
    A set $\X$ is said to be control forward invariant with respect to the system \eqref{eq:system} if for any $(x,w) \in \X \times \W$, there exists some control input $u \in \U$ such that $f(x, u) \in \X$. Now, if there exists a controller $g:\X \times \W \rightarrow \U$ such that the set $\X$ becomes control forward invariant, then the controller is said to be a forward invariant controller corresponding to the forward invariant set $\X$ with respect to the external input set $\W$.
\end{definition}

Now, we introduce the notion of $\delta$-ISS-CLF for the closed-loop system $\Xi_g$ where the sets $\X \subset \R^n, \W \subset \R^p$ are compact and $\X$ is considered to be control forward invariant under the controller $g$. Then, the definition of $\delta$-ISS-CLF becomes:

\begin{definition}\label{def:ISS-Lf}
     A smooth function $V:\X \times \X \rightarrow \R_0^+$ is said to be a $\delta$-ISS control Lyapunov function for closed-loop system $\Xi_g = (\X, \W, \U, f, g)$ in \eqref{eq:system}, where $\X, \W$ are compact sets and $g:\X \times \W \rightarrow \U$ be a forward invariant feedback controller, if there exist class $\mathcal{K}_\infty$ functions $\alpha_1, \alpha_2, \sigma$ and a constant $\kappa \in \R^+$ such that:
    \begin{enumerate}\label{cond:ISS-Lf}
        \item[(i)] for all  $x,\hat{x}\in \X$: $\alpha_1(|x-\hat{x}|) \leq V(x,\hat{x}) \leq \alpha_2(|x-\hat{x}|),$
        \item[(ii)] for all $x,\hat{x}\in \X$ and for all $w, \hat{w} \in \W: \frac{\partial V}{\partial x}f(x, g(x,w)) + \frac{\partial V}{\partial \hat{x}}f(\hat{x}, g(\hat{x},\hat{w})) \leq -\kappa V(x, \hat{x}) + \sigma(|w - \hat{w}|)$.
    \end{enumerate}
\end{definition} 

\begin{theorem}\label{th:admit}
The closed-loop ct-CS \eqref{eq:system} is said to be $\delta$-ISS within the compact state space $\X$ with respect to the external input $\w$, if there exists a $\delta$-ISS-CLF under the forward invariant controller $g$ as defined in Definition \ref{def:ISS-Lf}.
\end{theorem}
\begin{proof}
    The proof can be found in Appendix \ref{appendix:admit}.
\end{proof}

\subsection{Control Barrier Function}
To ensure that the controller $g$ makes the compact set $\X$ forward invariant, we leverage the notion of control barrier function as defined next. 
\begin{definition}[\cite{ames2019control}]\label{def:CBC}
    Given a ct-CS $\Xi=(\X,\U,f)$ with compact state and input sets $\X$ and $\U$. Let a $\mathcal{L}_{dh}$-smooth function, which is also Lipschitz continuous with Lipschitz constant $\mathcal{L}_h, h:\X \rightarrow \R$ is given as 
    \begin{subequations}\label{eq:leq_BC}
       \begin{align}
        h(x) &= 0, \hspace{0.3em} \forall x \in \partial \X, \\
        h(x) &> 0, \hspace{0.3em} \forall x \in int(\X).
    \end{align} 
    \end{subequations}
    Then, $h$ is said to be a control barrier function (CBF) for the system $\Xi$ in \eqref{eq:system} if for any $(x,w) \in \X \times \W$, there exists some control input $u \in \U$ such that the following condition holds:
    \begin{align} \label{eq:diff_BC}
        \frac{\partial h}{\partial x}f(x,u) \geq -\mu(h(x)), \quad \forall x \in \X.
    \end{align}
    where $\mu$ is a class $\mathcal{K}_{\infty}$ function.
\end{definition}

Now, based on Definition \ref{def:CBC}, we aim to design the forward invariant controller $u:=g(x,w)$ that will make the state-space $\X$ control forward invariant. The lemma below allows us to synthesize the controller, ensuring forward invariance. 

\begin{lemma}\label{lem:cfi_guarantee}
    The set $\X$ will be a control forward invariant set for system $\Xi$ in \eqref{eq:act_system} if there exists a function $h$ that satisfies the conditions of Definition \ref{def:CBC}.
\end{lemma}

\begin{proof}
    Consider a function $h: \X \rightarrow \R$ exists such that the conditions of \eqref{eq:leq_BC} holds,  \textit{i.e.}, $h(x) = 0, \forall x \in \partial \X$ and $h(x) > 0, \forall x \in int(\X)$.
    
    Now, we assume that there exists a controller $g:\X \rightarrow \U$ such that condition \eqref{eq:diff_BC} is satisfied. Then, one can infer at the boundary of $\X$, the derivative of $h$ is always positive, enforcing the trajectories of the system under the controller $g$ to move towards inside the state space. Hence, the state space $\X$ becomes control forward invariant. This completes the proof.
\end{proof}

\subsection{Problem Formulation}
Now, we formally define the main problem of this paper. 

\begin{problem}\label{prob}
    Given a continuous-time control system $\Xi = (\X, \U, f)$ as defined in \eqref{eq:act_system} with compact state-space $\X$ and unknown dynamics $f:\X \times \U \rightarrow \X$, the primary objective of the paper is to synthesize a forward invariant feedback controller $g: \X \times \W \rightarrow \U$ enforcing the closed-loop system $\Xi_g=(\X,\W,\U,f,g)$ in \eqref{eq:system} to be incrementally input-to-state stable with respect to external input $\w \in \W$ within the state space $\X$.
\end{problem}

One can see Problem \ref{prob} as finding the $\delta$-ISS-CLF that satisfies the conditions of Definition \ref{def:ISS-Lf} under the existence of some forward invariant controller $g$ for the given input space $\W$. 

In contrast to previous studies on controller design \cite{zamani2011backstepping}-\cite{zaker2024controller}, which depend on precise knowledge or a specific structure of the system dynamics, our objective is to develop a controller that achieves $\delta$-ISS for the closed-loop system without requiring exact knowledge or a defined structure of the dynamics.

To address the issues in determining the $\delta$-ISS-CLF and the corresponding controller, we present a neural network-based framework that satisfies the conditions of Definition \ref{def:ISS-Lf} and provides a formal guarantee for the obtained neural $\delta$-ISS-CLF and neural controller.

\section{Neural $\delta$-ISS Control Lyapunov Function}\label{sec:Neural Lyapunov}

To solve Problem \ref{prob}, we now try to design $\delta$-ISS-CLF, such that the closed-loop system becomes incrementally input-to-state stable according to Theorem \ref{th:admit}. To do so, we first provide the following lemma:
{
\begin{lemma}\label{lem:ROP}
    The closed-loop system in \eqref{eq:system} is assured to be $\delta$-ISS within the compact state-space $\X$ corresponding to the input-space $\W$ if there exists some $V$ and $g$ that satisfies the following conditions with $\eta \leq 0$:
    \begin{subequations} \label{eq:RCP}
    \begin{align} 
    & \forall x,\hat{x} \in \X, x = \hat{x}: \ V(x,\hat{x}) = 0, \\
    & \forall x,\hat{x} \in \X, x \neq \hat{x}: \ -V(x,\hat{x}) + \alpha_1(|x-\hat{x}|) \leq \eta, \label{eq:geq} \\
    &  \forall x,\hat{x} \in \X, x \neq \hat{x}: \ V(x,\hat{x}) - \alpha_2(|x-\hat{x}|) \leq \eta, \label{eq:leq} \\
    & \forall x,\hat{x} \in \X, x \neq \hat{x}, \forall w, \hat{w} \in \W: \notag \\
    & \quad \frac{\partial V}{\partial x}f(x, g(x,w)) + \frac{\partial V}{\partial \hat{x}}f(\hat{x}, g(\hat{x},\hat{w})) + \kappa V(x, \hat{x}) - \sigma(|w - \hat{w}|) \leq \eta ,\label{eq:diff} \\
    & \forall x \in \X,\forall w \in \W : \notag \\
    & \quad -\frac{\partial h}{\partial x}(f(x,g(x,w))) - \mu(h(x)) \leq \eta \label{eq:diff_BC_ROP}.
    \end{align}
    \end{subequations}
\end{lemma}
\begin{proof}
    With $\eta \le 0$, the condition \eqref{eq:geq} and \eqref{eq:leq} ensure that the $\delta$-ISS-CLF is bounded by the class $\mathcal{K}_\infty$ functions. Satisfaction of the condition \eqref{eq:diff} with $\eta \le 0$ enforces condition (ii) of Definition \ref{def:ISS-Lf} is satisfied. Finally, with $\eta \le 0$, satisfaction of condition \eqref{eq:diff_BC_ROP} implies the forward invariance of the set $\X$ under the controller $g$. Thereby, with $\eta \le 0$, satisfaction of the condition \eqref{eq:RCP} ensures $V$ to be a valid $\delta$-ISS-CLF.
\end{proof}

Note that $V(\cdot, \cdot) \in \{V|V: \X \times \X \rightarrow\R_0^+\}, g(\cdot, \cdot) \in \{g|g: \X \times \W \rightarrow\U\}, \alpha_1, \alpha_2, \mu \in \mathcal{K}_{\infty}, \sigma \in \mathcal{K}, {\kappa \in \R^+}$.
However, certain challenges prevent us to synthesize $V$ and $g$ using this lemma. They are listed as follows:
\begin{enumerate}
    \item[(C1)] The structures of the controller and the $\delta$-ISS-CLF are unknown. 
    \item[(C2)] The system dynamics $f$ is unknown; therefore, we cannot directly leverage the conditions \eqref{eq:diff} and \eqref{eq:diff_BC_ROP}.
    \item[(C3)] The structures of the $\mathcal{K}_{\infty}$ functions $\alpha_1, \alpha_2, \sigma, \mu$ and the constant $\kappa$ are unknown.
    \item[(C4)] There will be infinitely many constraints in the above lemma due to the continuous state space. 
\end{enumerate}}

To overcome the aforementioned challenges, the following subsections made some assumptions. To address challenge (C1), we parametrize $\delta$-ISS-CLF and the controller as feed-forward neural networks denoted by $V_{\theta,b}$ and $g_{\Bar{\theta}, \Bar{b}}$, respectively, where $\theta, \Bar{\theta}$ are weight matrices and $b, \Bar{b}$ are bias vectors. The detailed structures of these neural networks are discussed in the next subsection.

\subsection{Architecture of neural networks}
For a ct-CS $\Xi$ as in \eqref{eq:act_system}, the $\delta$-ISS-CLF neural network consists of an input layer with $2n$ neurons, where $n$ is the system dimension, and an output layer with one neuron, signifying the scalar output of the Lyapunov function. The network consists of $l_v$ hidden layers, with each hidden layer containing $h_v^i, i \in [1;l_v]$ neurons, where both values are arbitrarily chosen. To satisfy the condition (ii) of Definition \ref{def:ISS-Lf}, it requires the computation of $\frac{\partial V}{\partial x}$, which necessitates a smooth activation function $\varphi(\cdot)$ (for example, Tanh, Sigmoid, Softplus, etc.). Hence, the resulting Neural network function is obtained by applying the activation function recursively and is denoted by:
\[
\begin{cases}
t^0 = [x^\top,\hat{x}^\top]^\top , x,\hat{x} \in \R^n,  \\
t^{i+1} = \phi_i(\theta^it^i + b^i) \hspace{0.2 em} \text{for} \hspace{0.2 em} i \in [0;l_v-1], \\
V_{\theta,b}(x,\hat{x}) = \theta^{l_v}t^{l_v} + b^{l_v}, 
\end{cases}
\]
where $\phi_i:\R^{h_v^i} \rightarrow \R^{h_v^i}$ is defined as $ \phi_i(q^i) = [\varphi(q_1^i), \ldots, \varphi(q_{h_v^i}^i)]^\top$ with $q^i$ denoting the concatenation of outputs $q_j^i, j \in [1;h_v^i]$ of the neurons in $i$-th layer. The notation for the controller $g_{\Bar{\theta},\Bar{b}}$ is similar and is also considered a feed-forward neural network. In this case, the input layer and the output layer have dimensions of $n+p$ and $m$, respectively. The number of hidden layers of the controller neural network is $l_c$ and each layer has $h_c^i, i\in [1;l_c]$ neurons.

\subsection{Formal Verification Procedure of $\delta$-ISS-CLF}

Moving onto the challenge (C2), we make the following assumption.
\begin{assumption}\label{assum:black_box}
    {We consider having access to the black box or simulator model of the actual system \eqref{eq:act_system}. Hence, given an initial state $x$ and input signal $\mathsf{u}$, we will be able to forward simulate the system trajectory at any time $t \in \R_0^+$.}
\end{assumption}

This assumption allows us to use conditions \eqref{eq:diff} and \eqref{eq:diff_BC_ROP} without requiring explicit system knowledge, which is described at a later stage. Now, to overcome the challenge (C3), another assumption is considered.

\begin{assumption}\label{assum:K_infty}
    We consider that class $\mathcal{K}_\infty$ functions $\alpha_i, i \in \{1,2\}$ are of degree $\gamma_i$ with respect to $|x - \hat{x}|$ and class $\mathcal{K}_\infty$ function $\sigma$ is of degree $\gamma_w$ with respect to $|w - \hat{w}|$, i.e., $\alpha_i(|x - \hat{x}|)=k_i|x - \hat{x}|^{\gamma_i}$, and $\sigma(|w - \hat{w}|) = k_w|w - \hat{w}|^{\gamma_w}$, where $k:=[k_1, k_2, k_w], \Gamma:=[\gamma_1, \gamma_2, \gamma_w]$ are user-specific constants, where $\Gamma$ is chosen to make these functions convex.  Additionally, the class $\mathcal{K}_\infty$ function $\mu$ in \eqref{eq:diff_BC_ROP} is considered to be of the form $\mu_hh(x), \mu_h \in \R^+$, while the constants $\kappa, \mu_h \in \R^+$ are also user-defined.
\end{assumption}

Now, to avoid the infinite number of equations in the Lemma \ref{lem:ROP}, we leverage a sampling-based approach to get samples from the compact state space $\X$ as well as the input space $\W$. Note that these datasets will also help with the training procedure of the neural networks, which will be discussed in the next section. 
{Specifically, we draw $N$ state samples $x_s$ from the state space $\X$ where $s \in [1;N]$ and define a ball $B_{\varepsilon_x}(x_s)$ of radius $\varepsilon_x$ such that for all $x \in \X$, there exists $x_s$ such that $|x - x_s|\leq \varepsilon_x$. This ensures that $\bigcup_{s=1}^{N} B_{\varepsilon_x}(x_s) \supset \X$. Similarly, we sample $M$ inputs $w_p\in\W$ with balls of radius $\varepsilon_u$.}
We consider $\varepsilon = \max(\varepsilon_x, \varepsilon_u)$.  Collecting the data points obtained upon sampling the state-space $\X$ and input space $\W$, we form the training datasets denoted by:
\begin{align}\label{set:SCP}
\mathcal{X} \!=\! \{x_s | \!\bigcup_{s=1}^{N} \! B_{\varepsilon_x}(x_s) \!\supset\! \X \}, 
\Wo \!=\! \{w_p | \!\bigcup_{p=1}^{M}\! B_{\varepsilon_u}(w_p) \!\supset\! \W \}.
\end{align}
Now, we introduce the following assumptions concerning Lipschitz continuity to formulate the main theorem of this subsection.
\begin{assumption}\label{assum:Lipschitz_fun}
    The function $f$ in \eqref{eq:system} is Lipschitz continuous with respect to $x$ and $u$ with Lipschitz constants $\mathcal{L}_x$ and $\mathcal{L}_u$. {We assume the Lipschitz constants of the dynamics are known, even if the system is unknown.} 
\end{assumption}
Note that these constants can also be estimated using the approach in \cite[Algorithm 2]{FV_DD}.
\begin{assumption}\label{assum:Lipschitz_net}
   The candidate $\delta$-ISS-CLF is assumed to be Lipschitz continuous with Lipschitz bound $\mathcal{L}_L$ alongside its derivative is bounded by $\mathcal{L}_{dL}$. Similarly, the controller neural network has a Lipschitz bound $\mathcal{L}_C$. In the next section, we will explain how $\mathcal{L}_L, \mathcal{L}_{dL}$, and $\mathcal{L}_C$ are to be ensured in the training procedure. 
\end{assumption}
\begin{remark}
    Since, the sets $\X$ and $\W$ are compact, the class $\mathcal{K}_\infty$ functions are Lipschitz continuous with Lipschitz constants $\mathsf{L}_1, \mathsf{L}_2$ and $\mathsf{L}_u$, respectively, with respect to $|x-\hat{x}|$ and $|w-\hat{w}|$. The values can be estimated using the values of $k$ and $\gamma$. In addition, the Lipschitz constant $\mathcal{L}_h$ of the function $h$ is already known as the function $h$ is predefined.
\end{remark}
\begin{assumption}\label{assum:bound}
    We assume that the bounds of $\frac{\partial V}{\partial x}, \frac{\partial h}{\partial x}$, and $f(x,u)$ are given by $\mathcal{M}_L, \mathcal{M}_h$, and $\mathcal{M}_f$, respectively, \textit{i.e.,} $\sup_x|\frac{\partial V}{\partial x}| \leq \mathcal{M}_L, \sup_x|\frac{\partial h}{\partial x}| \leq \mathcal{M}_h$, and $\sup_{(x,u)}|f(x,u)| \leq \mathcal{M}_f$. 
\end{assumption}
{We first define the notation $\mathsf{L}_{f,g}V(x, \hat{x}):= \frac{\partial V_{\theta,b}}{\partial x}f(x, g_{\bar{\theta},\bar{b}}(x,w)) + \frac{\partial V_{\theta,b}}{\partial \hat{x}}f(\hat{x}, g_{\bar{\theta},\bar{b}}(\hat{x}, \hat{w}))$. However, since we have only a black-box model (as stated in Assumption \ref{assum:black_box}) to simulate the forward trajectory, we can not determine the actual value of $f(x, g_{\bar{\theta},\bar{b}}(x,w))$ for any $x \in \X, w \in \W$. To tackle the issue, we approximate the quantity as
\begin{align}\label{eq:lie_approx}
    \widehat{\mathsf{L}}_{f,g}V(x, \hat{x})&:= \frac{V_{\theta,b}\big(\mathsf{x}_{x,\mathsf{w}}(\tau),\mathsf{x}_{\hat{x},\hat{\mathsf{w}}}(\tau)\big) - V_{\theta,b}(x, \hat{x})}{\tau}, \   \forall \ x, \hat{x} \in \X, \forall \ \mathsf{w}(\tau), \hat{\mathsf{w}}(\tau) \in \W.
\end{align}
Now the proposed approximation \eqref{eq:lie_approx} satisfies the inequality $|\widehat{\mathsf{L}}_{f,g} V(x, \hat{x}) - \mathsf{L}_{f,g}V(x, \hat{x})| \leq \delta_V$, where $\delta_V \in \R^+$. To formally quantify the value of $\delta_V$, we first provide the following lemma.
\begin{lemma}\label{lem:Lipschitz_lie}
    Under Assumptions \ref{assum:Lipschitz_fun}, \ref{assum:Lipschitz_net} and \ref{assum:bound}, $\mathsf{L}_{f,g}V(x, \hat{x})$ is Lipschitz continuous with Lipschitz constant $\mathscr{L}_{Vx}$ in $x$ and $\mathscr{L}_{Vu}$ in $w$ where $\mathscr{L}_x := 2\mathcal{M}_L(\mathcal{L}_x + \mathcal{L}_u\mathcal{L}_c) + 2\mathcal{M}_f\mathcal{L}_{dL}$ and $\mathscr{L}_u:=2\mathcal{M}_L\mathcal{L}_u\mathcal{L}_c$.
\end{lemma}
\begin{proof}
    The proof can be found in Appendix \ref{appendix:Lipschitz_lie}.
\end{proof}

Now, the following theorem formally quantifies the closeness between $\mathsf{L}_{f,g}V(x, \hat{x})$ and its approximation $\widehat{\mathsf{L}}_{f,g}V(x, \hat{x})$.
\begin{theorem}\label{th:delta_lie}
    Let $\mathsf{L}_{f,g}V(x, \hat{x})$ be the derivative of the Lyapunov function over the pair of trajectories starting from the initial conditions $x$ and $\hat{x}$ while $\widehat{\mathsf{L}}_{f,g}V(x, \hat{x})$ be its approximation as in \eqref{eq:lie_approx}. Under Assumptions \ref{assum:Lipschitz_net}, \ref{assum:bound} and Lemma \ref{lem:Lipschitz_lie}, one has 
    \begin{align*}
        |\widehat{\mathsf{L}}_{f,g}V(x, \hat{x}) - \mathsf{L}_{f,g}V(x, \hat{x})| \leq \tau\mathscr{L}_{Vx}\mathcal{M}_f,
    \end{align*}
    where $\tau$ is the sampling time.
\end{theorem}
\begin{proof}
    The proof can be found in Appendix \ref{appendix:derivative}.
\end{proof}

Now one can leverage a similar notion to approximate the quantity $\mathsf{L}_{f,g}h(x_q):= \frac{\partial h}{\partial x_q}f(x_q,g_{\bar{\theta},\bar{b}}(x_q,w_q))$ as $\widehat{\mathsf{L}}_{f,g}h(x_q)$ satisfying the relation $|\widehat{\mathsf{L}}_{f,g}h(x_q) - \mathsf{L}_{f,g}h(x_q)| \leq \delta_h$, where the quantifier $\delta_h \in \R^+$ can be estimated via similar procedure as in Theorem \ref{th:delta_lie}. In addition, it is easy to show that $\mathsf{L}_{f,g}h(x_q)$ is Lipschitz continuous with respect to $x$ and $u$ with Lipschitz constants $\mathscr{L}_{hx}:=\mathcal{M}_f\mathcal{L}_{dh} + \mathcal{M}_h (\mathcal{L}_x + \mathcal{L}_u\mathcal{L}_C)$ and $\mathscr{L}_{hu}:=\mathcal{M}_h \mathcal{L}_u\mathcal{L}_C$ respectively.  Now we state the main theorem of this section.}
\begin{theorem}\label{th:constr}
    {The closed-loop system \eqref{eq:system} is guaranteed to be $\delta$-ISS within the compact state-space $\X$ if the following conditions are satisfied by the trained networks $V_{\theta,b}$ and $g_{\bar{\theta}, \bar{b}}$ with $\hat{\eta} + \mathcal{L}\varepsilon \leq 0$ where $\varepsilon=\max(\varepsilon_x, \varepsilon_u)$ as defined in \eqref{set:SCP} and $\mathcal{L} = \max\{\sqrt{2}\mathcal{L}_L + 2\mathsf{L}_1, \sqrt{2}\mathcal{L}_L + 2\mathsf{L}_2, \sqrt{2}\kappa\mathcal{L}_L + 2 \mathsf{L}_u + \mathscr{L}_{Vx} + \mathscr{L}_{Vu},  \mathscr{L}_{hx}+ \mathscr{L}_{hu} + \mu_h\mathcal{L}_h\}$}:
    \begin{subequations} \label{eq:SCP}
    \begin{align}
    &\forall x_q,x_r \in \mathcal{X}, x_q = x_r: V_{\theta,b}(x_q,x_r) = 0, \\
    &\forall x_q,x_r \in \mathcal{X}, x_q \neq x_r: -V_{\theta,b}(x_q,x_r) + k_1|x_q-x_r|^{\gamma_1} \leq \hat{\eta}, \label{eq:geq_SOP} \\
    & \forall x_q,x_r \in \mathcal{X}, x_q \neq x_r: V_{\theta,b}(x_q,x_r) - k_2|x_q-x_r|^{\gamma_2} \leq \hat{\eta}, \\
    & {\forall x_q,x_r \in \mathcal{X}, x_q \neq x_r, \forall w_q, w_r \in {\mathcal{W}}:} \notag \\
    & {\quad \widehat{\mathsf{L}}_{f,g}V(x_q, x_r) + \kappa V_{\theta,b}(x_q, x_r) - k_w|w_q - w_r|^{\gamma_w} + \delta_V \leq \hat{\eta}, \label{eq:diff_SOP}} \\
    & {\forall x_q \in \mathcal{X}, w_q \in \Wo: -\widehat{\mathsf{L}}_{f,g}h(x_q) - \mu_hh(x_q) - \delta_h \leq \hat{\eta}. \label{eq:diff_BC_SOP}}
    \end{align}
    \end{subequations}
\end{theorem}
\begin{proof}
    The proof can be found in Appendix \ref{appendix:constr}.
\end{proof}


\section{Training of Formally Verified $\delta$-ISS-CLF and the Controller}\label{sec:Training}
In this section, we propose the training procedure to solve Problem \ref{prob} while providing formal guarantees on the learned $\delta$-ISS-CLF and the controller with unknown dynamics. The structure of this section is as follows. We first present the formulation of suitable loss functions to synthesize $\delta$-ISS-CLF to solve Problem \ref{prob} and then demonstrate the training process that ensures formal guarantees.

\subsection{Formulation of Loss Functions}
The formulation of suitable loss functions is important to train the $\delta$-ISS-CLF and the controller such that minimization of the loss function leads to the satisfaction of the conditions \eqref{eq:SCP} for the training data set \eqref{set:SCP}. Now, we consider the conditions as sub-loss functions mentioned below:
\begin{subequations}
    \begin{align}
        L_1(\psi) &= \sum_{x,\hat{x} \in \mathcal{X}}\max\big(0,(V_{\theta,b}(x,\hat{x}) \big), \\
        L_2(\psi) &= \sum_{x,\hat{x} \in \mathcal{X}}\max\big(0,(-V_{\theta,b}(x,\hat{x}) + k_1|x- \hat{x}|^{\gamma_1} - \hat{\eta})\big), \\
        L_3(\psi) &= \sum_{x,\hat{x} \in \mathcal{X}}\max\big(0,(V_{\theta,b}(x,\hat{x}) - k_2|x- \hat{x}|^{\gamma_2} - \hat{\eta})\big), \\
        L_4(\psi) &= {\hspace{-0.5cm} \sum_{x,\hat{x} \in \mathcal{X}, w, \hat{w} \in \Wo} \hspace{-0.5cm} \max\big(0,( \widehat{\mathsf{L}}_{f,g}V(x_q, x_r) + \kappa V_{\theta,b}(x, \hat{x})  - k_w|w - \hat{w}|^{\gamma_w} + \delta_V - \hat{\eta})\big), }\\
        L_5(\psi) &= {\hspace{-0.5cm} \sum_{x \in \mathcal{X}, w \in \Wo} \hspace{-0.4cm} \max\big(0, (-\widehat{\mathsf{L}}_{f,g}h(x_q) - \mu_hh(x_q) - \delta_h - \hat{\eta})\big)},
\end{align}
\end{subequations}
where $\psi = [\theta,b,\Bar{\theta},\Bar{b}]$ are trainable parameters. As mentioned, the actual loss function is a weighted sum of the sub-loss functions and is denoted by
\begin{equation}\label{eq:loss_LF}
    L(\psi) = \sum_{i=1}^5c_iL_i(\psi), 
\end{equation}
where $c_i \in \R^+, i \in [1;5]$ are the weights of the sub-loss functions $L_i(\psi)$.

To ensure Assumption \ref{assum:Lipschitz_net}, it is crucial to verify the Lipschitz boundedness of $V_{\theta,b}, \frac{\partial V_{\theta,b}}{\partial x}$ and $g_{\Bar{\theta},\Bar{b}}$ with corresponding Lipschitz bounds denoted by $\mathcal{L}_L, \mathcal{L}_{dL}$ and $\mathcal{L}_C$ respectively. To train the neural network with Lipschitz bounds, we have the following lemma.
\begin{lemma}[\cite{fazlyab2019efficient}] \label{lem:Lipschitz_NN}
    Suppose $f_\theta$ is a $\mathcal{N}$-layered feed-forward neural network with $\theta = [\theta_0, \ldots, \theta_{\mathcal{N}}]$ as the trainable parameter where $\theta_i, i \in [0;\mathcal{N]}$ is the weight of $i$-th layer. Then the neural network is said to be Lipschitz bounded with the Lipschitz constant $\mathcal{L}_L$ is ensured by the following semi-definite constraint $M_{\mathcal{L}_L}(\theta,\Lambda)$:=
    \begin{align}\label{eq:mat_ineq}
        \begin{bmatrix}
            A \\ B
        \end{bmatrix}^\top
        \begin{bmatrix}
            2 \alpha \beta \Lambda & -(\alpha+\beta)\Lambda \\ -(\alpha+\beta)\Lambda & 2\Lambda
        \end{bmatrix}
        \begin{bmatrix}
            A \\ B
        \end{bmatrix}
        + \notag \\
        \begin{bmatrix}
            \mathcal{L}_L^2\textbf{I} & 0 & 0 & 0 \\ 0 & 0 & 0 & 0 \\ 0 & 0 & 0 & -\theta_\mathcal{N}^\top \\ 0 & 0 & -\theta_\mathcal{N} & \textbf{I} 
        \end{bmatrix} \geq 0,
    \end{align}
    where 
    \begin{align*}
        A = \begin{bmatrix}
        \theta_0 & \ldots & 0 & 0 \\ \vdots & \ddots & \vdots & \vdots \\0 & \ldots & \theta_{\mathcal{N}-1} & 0  
    \end{bmatrix}, B = \begin{bmatrix}
        0 & \textbf{I}
    \end{bmatrix},
    \end{align*}
    where $\theta_0, \ldots, \theta_\mathcal{N}$ are the weights of the neural network, $\Lambda \in \mathcal{D}_{\geq 0}^{n_i}, i \in \{1, \ldots, \mathcal{N}\}$, where $n_i$ denotes number of neurons in $i$-th layer, and $\alpha,$ and $ \beta$ are the minimum and maximum slope of the activation functions, respectively. {It should be carefully noted that the dimensions of the matrices $A$ and $B$ are $(\sum_{i=1}^{\mathcal{N}} n_i ) \times (1+ \sum_{i=0}^{\mathcal{N}-1} n_i )$ and $(\sum_{i=1}^{\mathcal{N}} n_i ) \times (1+ \sum_{i=1}^{\mathcal{N}} n_i )$ respectively}
\end{lemma}

The lemma addresses the certification of the Lipschitz bound for a neural network, but our scenario also requires the Lipschitz boundedness of $\frac{\partial V_{\theta,b}}{\partial x}$. To address the issue, we make the following assumption.

\begin{assumption}\label{assum:lipschitz_act_fun}
    We consider the activation functions of the feedforward neural network $V_{\theta,b}$, denoted by $\phi_i, i \in \{1, \ldots, \No\}$ where $\No$ is the number of layers of the neural network, is Lipschitz bounded by the constants $\Lo_i, i \in \{1, \ldots, \No\}$, \textit{i.e.,} $\lVert \phi_i(x) - \phi_i(y) \rVert \leq \Lo_i \lVert x - y \rVert$ for all $x,y\in \X$. 
\end{assumption}

Now, based on the aforementioned assumption, we propose the following theorem in order to satisfy the Lipschitz continuity of the derivative of the neural network.

\begin{theorem}\label{th:derivative}
    Consider a $\No$-layered feed-forward neural network $f_\theta$, with the output being a scalar, where $\theta$ represents trainable weight and bias parameters. Let $y \in \R$ denote the scalar output of the neural network, $x \in \R^n$ denote the input of the neural network, $\theta_i, i \in \{0,1,\ldots,\No\}$ denote the weight parameters and $\phi_i, i \in \{1,\ldots,\No\}$ denote the activation functions of $i$-th layer. Then, the Lipschitz continuity of the derivative of the neural network $\frac{\partial y}{\partial x}$ can be ensured by $M_{\mathcal{L}_{dL}}(\hat{\theta}, \hat{\Lambda}) \geq 0$ using the Lemma \ref{lem:Lipschitz_NN}, where 
    \begin{subequations}
        \begin{align}
            \hat{\theta} &= (\theta_0, \theta_1, \ldots, \theta_{\No-1},\hat{\theta}_{\No}), \\
            \hat{\theta}_{\No} &= \Lo \theta_0^\top \theta_1^\top \ldots \theta_{\No-1}^\top\text{diag}(\theta_{\No})\label{last_layer}
        \end{align}
    \end{subequations}
    with $\Lo = \Lo_1 \ldots \Lo_{\No-1}$, $\hat{\Lambda} \in \mathcal{D}_{\geq 0}^{n_i}$, similar to $\Lambda$ as in Lemma \ref{lem:Lipschitz_NN} and $\mathcal{L}_{dL}$ being the Lipschitz bound of the derivative.
\end{theorem}
\begin{proof}
    The proof can be found in Appendix \ref{appendix:derivative}.
\end{proof}
\begin{remark}
    The proof shows that the derivative of a neural network with $\No$ layers and Lipschitz continuous activation functions is another neural network,
    where the change is mainly reflected in the weights of the last layer as given in \eqref{last_layer}.
\end{remark}
Now, to ensure the loss function satisfies the matrix inequalities corresponding to the Lipschitzness of Lyapunov, its derivative and the controller, we characterize another loss function denoted by,
\begin{align}\label{eq:loss_ineq}
    &L_M(\psi,\Lambda,\hat{\Lambda},\Bar{\Lambda}) = -c_{l_1}\log\det(M_{\mathcal{L}_L}(\theta,\Lambda)) \notag \\
    \quad & - c_{l_2}\log\det(M_{\mathcal{L}_{dL}}(\hat{\theta},\hat{\Lambda})) -c_{l_3}\log\det(M_{\mathcal{L}_C}(\Bar{\theta},\Bar{\Lambda})), 
\end{align}
where $c_{l_1}, c_{l_2}, c_{l_3} \in \R^+$ are weights for sub-loss functions, $M_{\mathcal{L}_L}(\theta,\Lambda), M_{\mathcal{L}_{dL}}(\hat{\theta},\hat{\Lambda})$, and $ M_{\mathcal{L}_C}(\Bar{\theta},\Bar{\Lambda})$ are the matrices corresponding to the bounds $\mathcal{L}_L, \mathcal{L}_{dL}$, and $\mathcal{L}_C$, respectively.

\subsection{Training with Formal Guarantee}
The training process of the neural Lyapunov function and the controller is described in Algorithm \ref{algo:NN_training}.

\begin{algorithm}[h]
\caption{Training of the Neural Networks}
\label{algo:NN_training}
{\begin{algorithmic}[1]
    \Require Black box model of the system $f(x,u)$, Data sets: $\mathcal{X}, \mathcal{W}$ 
    \Ensure $V_{\theta,b}, g_{\bar{\theta}, \bar{b}} $
    \State Select the hyperparameters $\epsilon$, $c = [c_0, c_1, c_2, c_3, c_{l1}, c_{l2}, c_{l3}], k = [k_1, k_2, k_w], \Upsilon = [\gamma_1, \gamma_2, \gamma_w], \kappa, \mu_h, \mathcal{L}_L, \mathcal{L}_C, \mathcal{L}_{dL}, \mathcal{L}_h$ and number of epochs $n_{ep}$.
    \State Computation of Lipschitz constants $\mathsf{L}_1, \mathsf{L}_2$ and $\mathsf{L}_u$. Compute $\mathcal{L}$ using Theorem \ref{th:constr}.
    \State Initialize Neural networks and trainable parameters $\theta,b,\bar{\theta}, \bar{b},\Lambda,\bar{\Lambda}$. Initialize $\hat{\eta}$ as $\hat{\eta} = - \mathcal{L}\varepsilon $.
    \For{$i\leq Epochs$ (Training starts here)}
        \State Create batches of training data from $\mathcal{X}, \mathcal{W}$
        \State Find batch losses using \eqref{eq:loss_LF} and \eqref{eq:loss_ineq}.
        \State Use ADAM or SGD optimizer with specified learning rate \cite{ruder2016} to reduce loss and update the trainable parameters.
        \State \textbf{If} Theorem \ref{th:guarantee} is satisfied $\rightarrow$ \textbf{break}.
    \EndFor
    \State \textbf{return} $V_{\theta,b}, g_{\bar{\theta}, \bar{b}}.$
\end{algorithmic}}
\end{algorithm}

We present the theorem that provides the formal guarantee for the incremental stable nature of the closed-loop system under the action of the controller $g_{\bar{\theta}, \bar{b}}$.
\begin{theorem}\label{th:guarantee}
    Consider a continuous-time control system $\Xi$ as in \eqref{eq:act_system} with compact state-space $\X$ and input space $\W$. Let, $V_{\theta,b}$ and $g_{\Bar{\theta},\Bar{b}}$ denote the trained neural networks representing the $\delta$-ISS control Lyapunov function and the controller such that $L(\psi, \eta) = 0$ and $L_M(\psi,\Lambda,\hat{\Lambda},\Bar{\Lambda}) \leq 0$ over the training data sets $\mathcal{X}$ and $\mathcal{W}$. Then, the closed-loop system under the influence of the controller $g_{\Bar{\theta},\Bar{b}}$ is guaranteed to be incrementally ISS within the state-space.
\end{theorem}
\begin{proof}
{The first loss $L(\psi,\eta)=0$ is defined as a weighted sum of its constituent sub-loss terms, therefore the condition $L(\psi,\eta)=0$ implies that its individual sub-loss components vanishes, which in turn, ensures the conditions of Theorem \ref{th:constr} has been solved with  $\hat{\eta}$. This enforces the satisfaction of all the inequalities of the $\delta$-ISS-CLF conditions corresponding to Definition \ref{def:ISS-Lf} for the sampled datapoints in $\mathcal{X}$ and $\mathcal{W}$ corresponding to \eqref{set:SCP}. Note that with the initialization of $\hat{\eta} = -\mathcal{L}\varepsilon$, the condition of Theorem \ref{th:constr} has already been satisfied, resulting closed-loop system to be $\delta$-ISS for any initial state in $\X$ and any external input $\W$.} Another loss $L_{\mathcal{M}}(\psi,\Lambda, \hat{\Lambda},\bar{\Lambda}) \leq 0$ ensures the predefined Lipschitz constants of the networks. Hence, the satisfaction of the above theorem leads to ensuring that the closed-loop system is $\delta$-ISS under the action of the controller.
\end{proof}
\begin{remark}
    {Note that if the algorithm fails to converge, the $\delta$-ISS of the closed-loop system cannot be determined for the given hyperparameters $c,k,\Gamma, \mathcal{L}_L,\mathcal{L}_C$. Convergence can be improved by reducing the discretization parameter $\epsilon$ \cite{zhao2021learning}, tuning neural network hyperparameters (e.g., architecture, learning rate) \cite{nn_lr}, or revising the algorithm’s initial hyperparameters.}
\end{remark}

\begin{remark}
    In addition, the initial feasibility of condition \eqref{eq:mat_ineq} is required to satisfy the criterion of loss $L_M$ in \eqref{eq:loss_ineq} according to Theorem \ref{th:guarantee}. Choosing small initial weights and bias of the neurons can ensure this condition \cite{pauli2022a}.
\end{remark}

\begin{remark}[Dealing with input constraints]
    To keep the output of the controller neural network bounded within the input constraints, one can consider the HardTanh function as the activation function of the last layer of the controller neural network. In that case, $\U$ is assumed to be a polytopic set with bounds given as $u_{\min} \leq u \leq u_{\max}, u \in \U$. Then, the resulting controller will be:
    $\begin{cases}
    z^0 = [x^\top,w^\top]^\top , x \in \X, w \in \W, \\
    z^{i+1} = \phi_i(\bar{\theta}^iz^i + \bar{b}^i) \hspace{0.2 em} \text{for} \hspace{0.2 em} i \in [0;l_c-1], \\
    g_{\Bar{\theta},\Bar{b}}(x,w) =
    \begin{cases}
        u_{\min}, \quad \bar{\theta}^{l_c}z^{l_c} + \bar{b}^{l_c} \leq u_{\min}, \\
        u_{\max}, \quad \bar{\theta}^{l_c}z^{l_c} + \bar{b}^{l_c} \geq u_{\max}, \\
        \bar{\theta}^{l_c}z^{l_c} + \bar{b}^{l_c}, \quad \text{otherwise}.
    \end{cases}
    \end{cases}$
\end{remark}

\section{Case Study}\label{sec:Cases}
The proposed method for achieving incremental input-to-state stability in a system using a neural network-based controller is demonstrated through multiple case studies. All the case studies were performed using PyTorch in Python 3.10 on a machine with a Windows operating system with an Intel Core i7-14700 CPU, 32 GB RAM and NVIDIA GeForce RTX 3080 Ti GPU.

\subsection{Scalar Nonaffine Nonlinear System} \label{Case study 1}
For the first case study, we consider a scalar nonaffine nonlinear system, whose dynamics is given by:
\begin{align}
    \dot{\mathsf{x}}(t) = a \big(\sin(\mathsf{x}(t)) + \tan(\mathsf{u}(t))\big),
\end{align}
where $\mathsf{x}(t)$ denotes the state of the system at time instant $t$. The constant $a$ represents the rate constant of the system. We consider the state space of the system to be $\X = [-\frac{\pi}{2}, \frac{\pi}{2}]$. Moreover, the external input set is bounded within $\W = [-0.5,0.5]$. We consider the model to be unknown. However, we estimate the Lipschitz constants $\mathcal{L}_x = 0.2, \mathcal{L}_u = 0.25$ using the results in \cite{Lipschitz}. 

The goal is to synthesize a controller to enforce the closed-loop system to be $\delta$-ISS. So, we will try to synthesize a valid $\delta$-ISS-CLF $V_{\theta,b}$ under the action of the controller $g_{\Bar{\theta},\Bar{b}}$. To do this, we first fix the training hyper-parameters as $\epsilon = 0.0016, \mathcal{L}_L = 1, \mathcal{L}_{dL} = 1, \mathcal{L}_C = 5, k_1 = 0.00001, k_2 = 1, k_w = 0.01, \kappa = 0.0001, \mu_h = 0.0001$. So, the overall Lipschitz constant according to Theorem \ref{th:constr} is $3.9555$. We fix the structure of $V_{\theta,b}$ as $l_f = 1, h_f^1 = 40$ and $g_{\Bar{\theta},\Bar{b}}$ as $l_c = 1, h_c^1 = 15$. The activation functions for $\delta$-ISS-CLF and the controller are Tanh and ReLU functions, respectively.

Now, we consider the training data obtained from \eqref{set:SCP} and perform training to minimize the loss functions $L, L_\mathcal{P} $ and $ L_v$. The training algorithm converges to obtain the $\delta$-ISS-CLF $V_{\theta,b}$ along with $\eta = -0.0065$. Hence, $\eta+\mathcal{L}\epsilon = -0.0065 + 3.9555\times0.0016 = -0.00017$, thus, using Theorem \ref{th:guarantee}, we can guarantee that the obtained Lyapunov function $V_{\theta,b}$ is valid and the closed-loop system is assured to be incrementally input-to-state stable under the influence of the controller $g_{\Bar{\theta},\Bar{b}}$.

The successful runs of the algorithm have an average convergence time of 30 minutes.

\begin{figure*}[t]
    \centering
    \begin{subfigure}{0.32\textwidth}
        \centering
        \includegraphics[width=\textwidth]{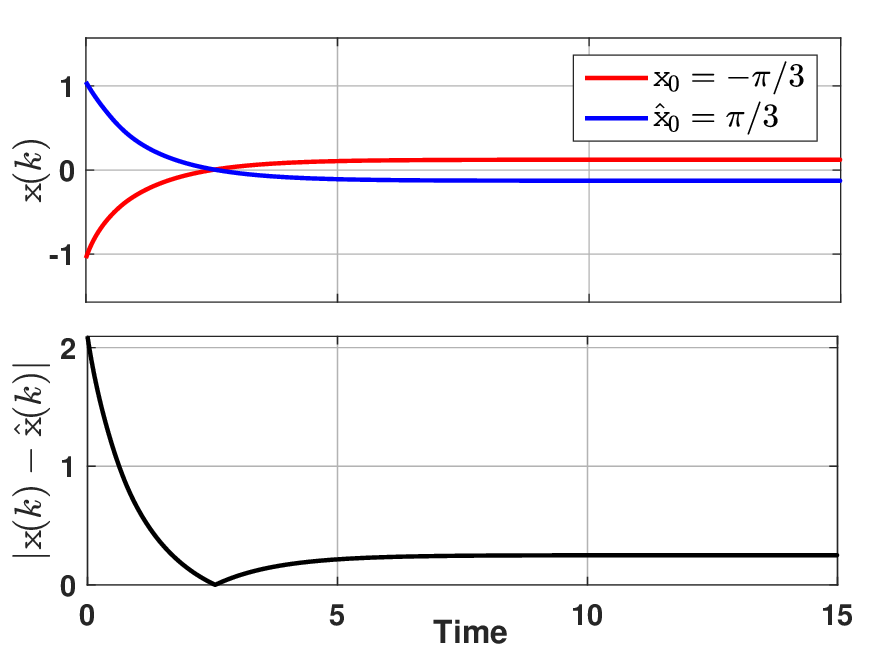}
        \caption{}
    \end{subfigure}
    \hfill
    \begin{subfigure}{0.32\textwidth}
        \centering
        \includegraphics[width=\textwidth]{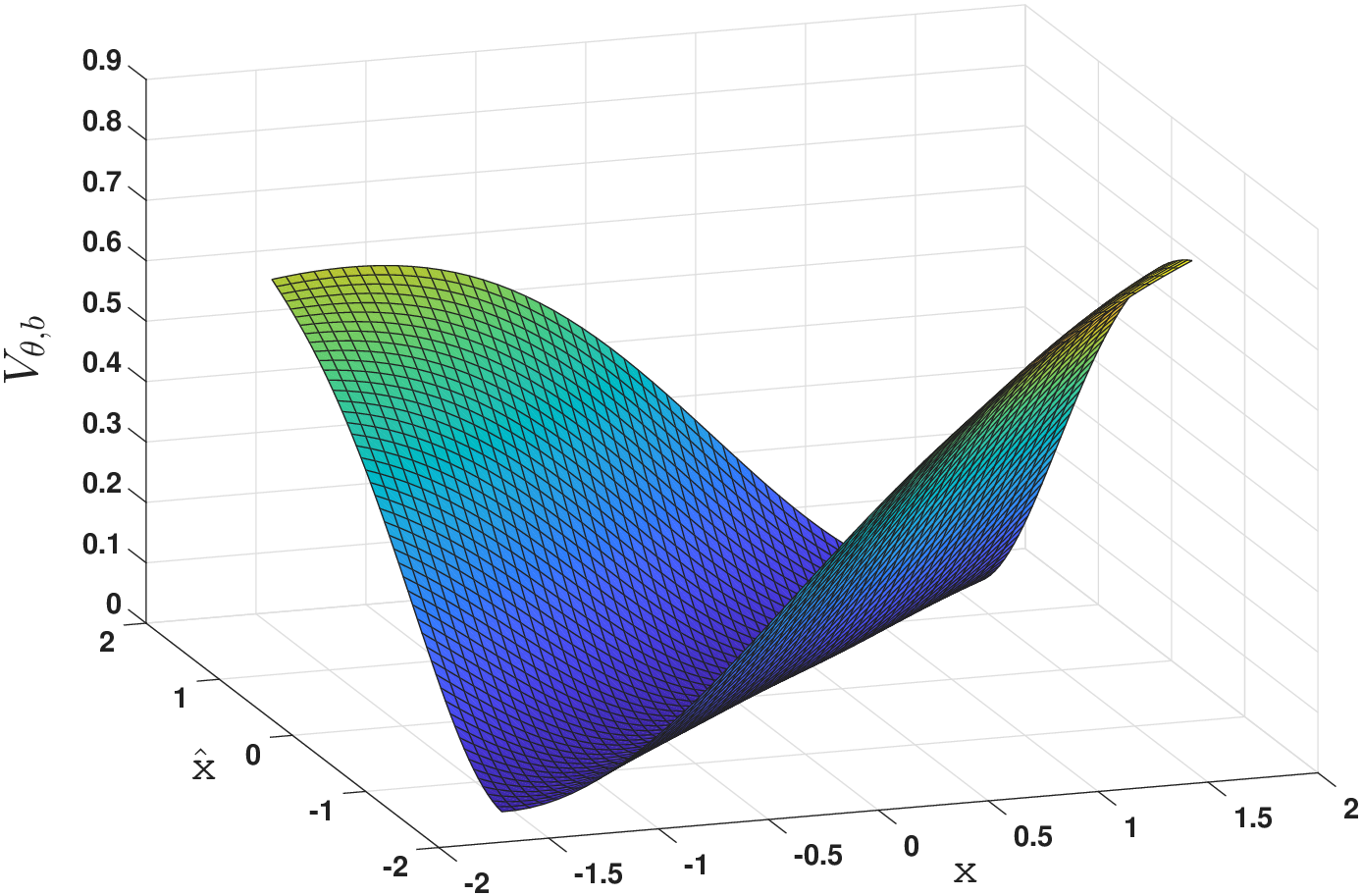}
        \caption{}
    \end{subfigure}
    \hfill
    \begin{subfigure}{0.32\textwidth}
        \centering
        \includegraphics[width=\textwidth]{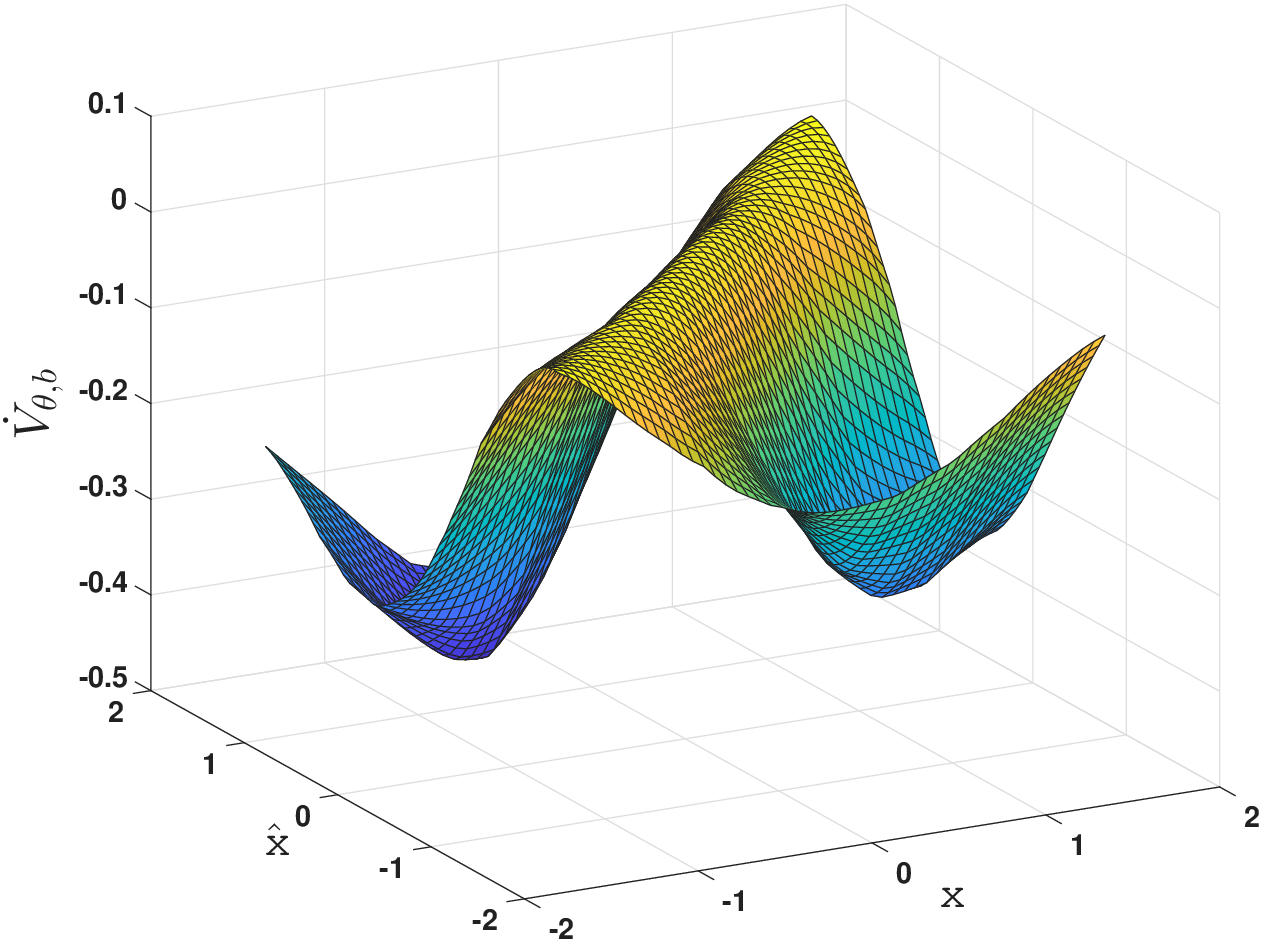}
        \caption{}
    \end{subfigure}
    \vspace{-0.2cm}
    \caption{(a)Trajectories starting from different initial conditions under different input signals (for all $t \geq 0, \w(t) = -0.1 \in \W$ for blue curve and $\w(t) = 0.1 \in \W$ for red curve), (b) The $\delta$-ISS-CLF plot is greater than zero for all $(x, \hat{x}) \in \X \times \X$, (c) The derivative of $V$ is always negative corresponding to the same input $\w = \hat{\w}$ satisfying the condition \eqref{eq:diff}.}
    \label{fig:sim1}
\end{figure*}

As can be seen in Figure \ref{fig:sim1}(a), under the different input conditions, the trajectories starting from different initial conditions maintain the same distance after some time instances under the influence of the controller. The $\delta$-ISS-CLF plot for this case is shown in Figure \ref{fig:sim1}(b). Note that in Figure \ref{fig:sim1}(c), the plot of condition \eqref{eq:diff} is shown under the same input signal, and it is always negative, satisfying the condition.

\subsection{Single-link Manipulator}

We consider a single link manipulator dynamics \cite{lewis2020neural,murray2017mathematical}, whose dynamics is governed by the following set of equations.
\begin{align}
    \dot{\mathsf{x}}_1(t) &= \mathsf{x}_2(t),\notag \\
    \dot{\mathsf{x}}_2(t) &= \frac{1}{M}(\mathsf{u}(t) - b\mathsf{x}_2(t)),
\end{align}
where $\mathsf{x}_1(t),\mathsf{x}_2(t)$ denotes angular position and velocity at time instance $t$ respectively. The constants $M = 1, b = 0.1$ represent the mass and damping coefficient of the system, respectively. We consider the state space of the system to be $\X = [-\frac{\pi}{6}, \frac{\pi}{6}]\times[-\frac{\pi}{6}, \frac{\pi}{6}]$. Moreover, we consider the input set to be bounded within $\W = [-0.5,0.5]$. Also, we consider the model to be unknown. However, we estimate the Lipschitz constants $\mathcal{L}_x = 1.005, \mathcal{L}_u =1$. 

The goal is to synthesize a controller to enforce the system to be $\delta$-ISS. To do this, we first fix the training hyper-parameters as $\epsilon = 0.0105, \mathcal{L}_L = 1, \mathcal{L}_{dL} = 1, \mathcal{L}_C = 2, k_1 = 0.0001, k_2 = 1, k_u = 0.001, \mu_h = 0.0001, \kappa = 0.00001$. So, the Lipschitz constant according to Theorem \ref{th:constr} is $7.6689$. We fix the structure of $V_{\theta,b}$ as $l_f = 1, h_f^1 = 60$ and $g_{\Bar{\theta},\Bar{b}}$ as $l_c = 1, h_c^1 = 20$. 
The activation functions for $\delta$-ISS-CLF and the controller are Softplus and ReLU, respectively. 
The training algorithm converges to obtain the $\delta$-ISS-CLF $V_{\theta,b}$ along with $\eta = -0.0806$. Hence, $\eta+\mathcal{L}\epsilon = -0.0806 + 7.6689\times0.0105 = -0.000076$, thus by utilizing Theorem \ref{th:guarantee}, we can guarantee the obtained $\delta$-ISS-CLF $V_{\theta,b}$ is valid and the closed-loop system is assured to be incrementally input-to-state stable under the influence of the controller $g_{\Bar{\theta},\Bar{b}}$.

The successful runs of the algorithm have an average convergence time of 2 hours.

\begin{figure}[h]
    \centering
    \includegraphics[width=0.6\linewidth]{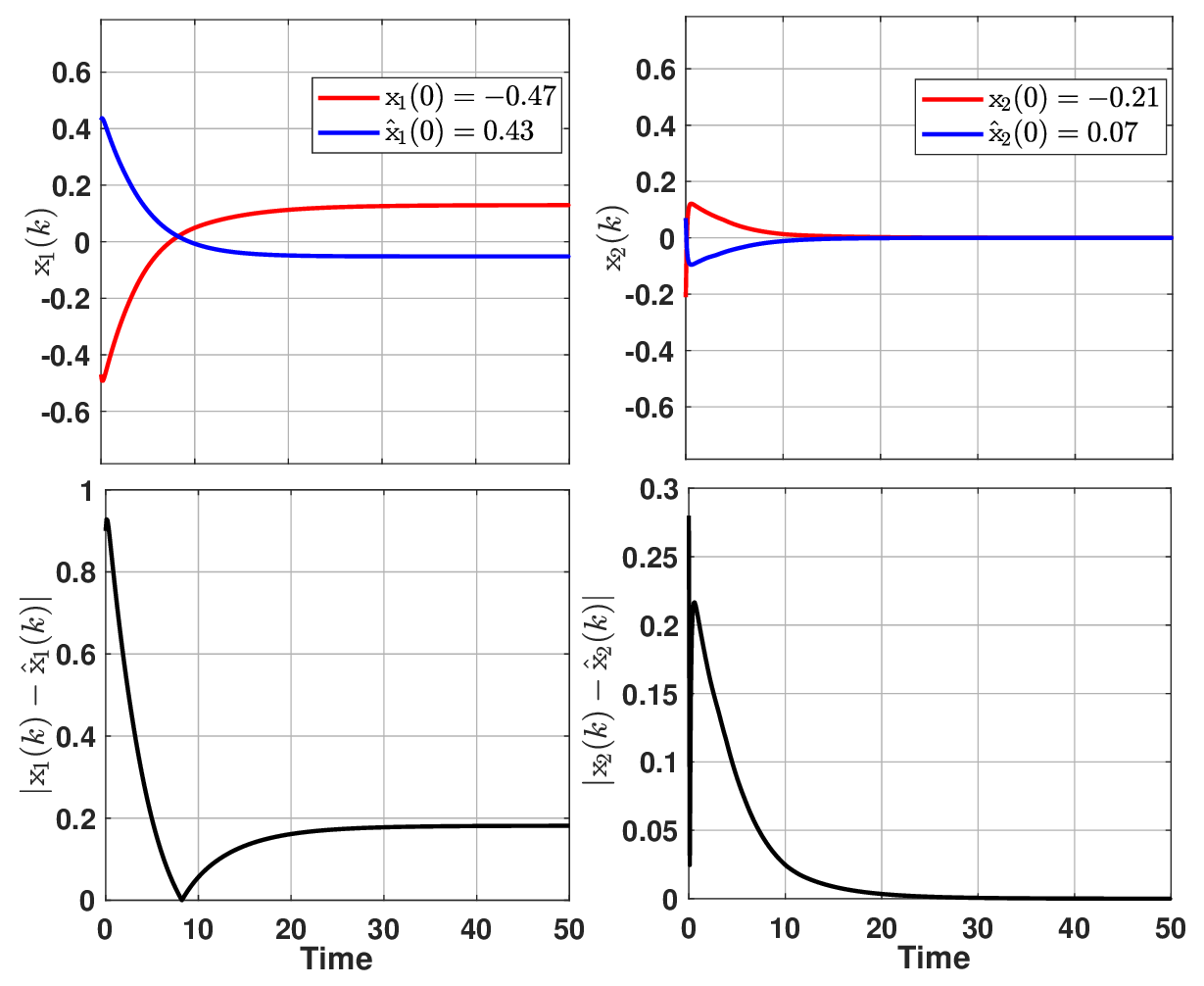}
    \caption{Top: Angular position (left) and velocity (right) of the manipulator, where the blue curve is influenced under input $\w(t) = -0.1 \in \W$, and the red curve is influenced under input $\w(t) = 0.2 \in \W$ for all $t \geq 0$. Bottom: The difference in angular positions (left) and velocities (right) subjected to different initial conditions and input torques.}
    \label{fig:sim2}
\end{figure}

One can see from Figure \ref{fig:sim2} that under the different input conditions, the trajectories corresponding to various states starting from different initial conditions maintain the same distance or converge to a particular trajectory after some time instances under the influence of the controller.

\subsection{Magnetic Levitator System}
For the next case study, we consider a magnetic levitator system \cite{ghosh2014design}, whose dynamics is given by:
\begin{align}
\dot{\mathsf{x}}_1(t) &= \mathsf{x}_2(t), \notag\\
    \dot{\mathsf{x}}_2(t) &= g - \frac{k}{m} \frac{i^2}{\mathsf{x}_1^2(t)}
\end{align}
where $\mathsf{x}(t) = [\mathsf{x}_1(t), \mathsf{x}_2(t)]$ denotes the position and velocity of the ball at time instant $t$, $m = 0.01$ stands for the mass of the ball, $g=9.81$ being the gravitational constant and $k=0.009$ being the coil constant of the system. The input to the system is the coil current $i$. We consider the state space to be $\X = [0.25, 0.75] \times [-0.25,0.25]$. For this case study, we show that the controller is able to make the system $\delta$-GAS, as there is no external input. We estimate the Lipschitz constants to be $\mathcal{L}_x = 1, \mathcal{L}_u=0.15$.  

We first fix the training hyper-parameters as $\epsilon = 0.005, \mathcal{L}_L = 1, \mathcal{L}_{dL} = 1, \mathcal{L}_C = 2, k = [0.00001, 0.2, 0.01], \kappa = 0.00001, \mu_h = 0.1$. So, the Lipschitz constant according to Theorem \ref{th:constr} is $2.8685$.
We fix the structure of $V_{\theta,b}$ as $l_f = 2, h_f^1 = 60$ and $g_{\Bar{\theta},\Bar{b}}$ as $l_c = 1, h_c^1 = 20$. The activation functions for $\delta$-ISS-CLF and the controller are Softplus and ReLU, respectively.
The training algorithm converges to obtain $\delta$-ISS-CLF $V_{\theta,b}$ along with $\eta = -0.0162$. Hence, $\eta+\mathcal{L}\epsilon = -0.0162 + 2.8685\times0.005 = --0.00186$, therefore, using Theorem \ref{th:guarantee}, we can guarantee that the obtained $\delta$-ISS-CLF $V_{\theta,b}$ is valid and that the closed-loop system is assured to be incrementally input-to-state stable under the influence of the controller $g_{\Bar{\theta},\Bar{b}}$.

The successful runs of the algorithm have an average convergence time of 2 hours.

\begin{figure}[h]
    \centering
    \includegraphics[width=0.6\linewidth]{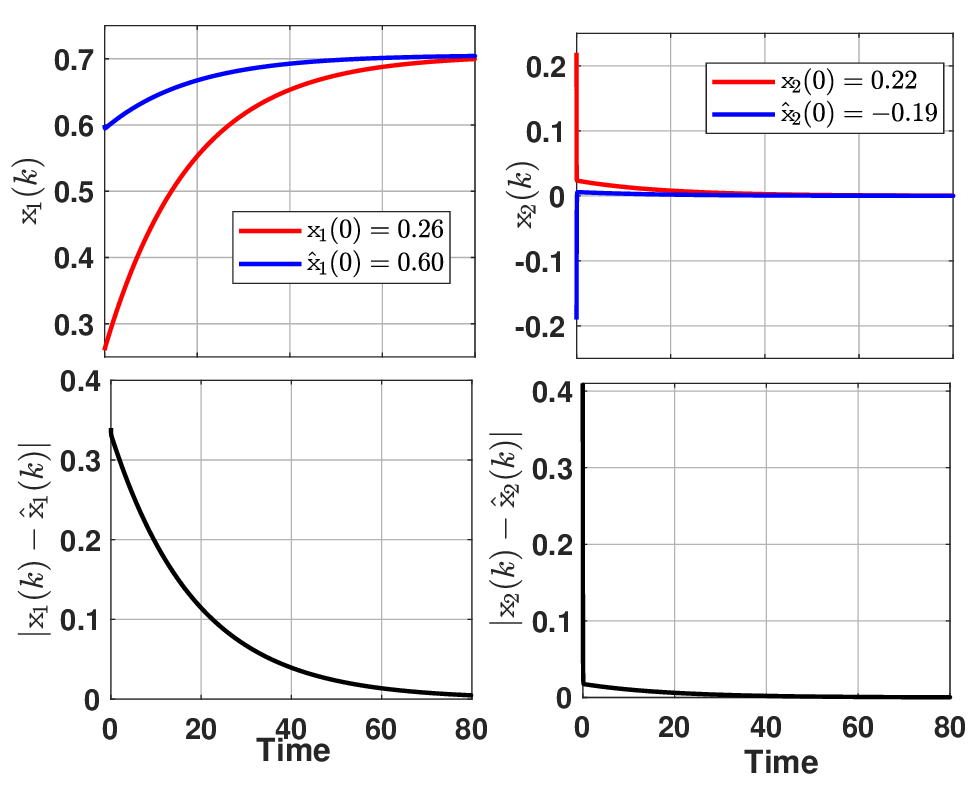}
    \caption{Top: Position (left) and velocity (right) of the ball, where the trajectories converge towards each other without the presence of external input for all $t \geq 0$. Bottom: The difference in positions (left) and velocities (right) subjected to different initial conditions and input currents.}
    \label{fig:sim5}
\end{figure}

One can see from Figure \ref{fig:sim5} that under the different input conditions, the trajectories corresponding to various states starting from different initial conditions converge to a particular trajectory after some time instances under the influence of the controller.

\subsection{Jet Engine}
We consider a nonlinear Moore-Grietzer Jet Engine Model in no-stall mode \cite{jet_engine}, whose dynamics is governed by the following set of equations.
\begin{align}
    \dot{\mathsf{x}}_1(t) &= - \mathsf{x}_2(t) - 1.5\mathsf{x}_1^2(t) - 0.5\mathsf{x}_1^3(t),\notag \\
    \dot{\mathsf{x}}_2(t) &= \mathsf{u}(t),
\end{align}
where $\mathsf{x}_1 = \mu - 1,\mathsf{x}_2 = \zeta - \rho - 2$ with $\mu, \zeta, \rho$ denote the mass flow, the pressure rise and a constant respectively. We consider the state space of the system to be $\X = [-0.25,0.25]\times[-0.25,0.25]$. Moreover, we consider the input set to be bounded within $\W = [-0.25,0.25]$. Also, we consider the model to be unknown. However, we estimate the Lipschitz constants $\mathcal{L}_x = 1.213, \mathcal{L}_u = 1$.

The goal is to synthesize a controller to enforce the system to be $\delta$-ISS. To do this, we first fix the training hyper-parameters as $\epsilon = 0.005, \mathcal{L}_L = 1, \mathcal{L}_{dL} = 1, \mathcal{L}_C = 2, k_1 = 0.00001, k_2 = 2, k_w = 0.01, \kappa = 0.0001, \mu_h = 1$. So, the Lipschitz constant according to Theorem \ref{th:constr} is $8.103$.
We fix the structure of $V_{\theta,b}$ as $l_f = 1, h_f^1 = 60$ and $g_{\Bar{\theta},\Bar{b}}$ as $l_c = 1, h_c^1 = 20$. The activation functions for $\delta$-ISS-CLF and the controller are Softplus and ReLU, respectively.
The training algorithm converges to obtain the $\delta$-ISS-CLF $V_{\theta,b}$ along with $\eta = -0.0410$. Hence, $\eta+\mathcal{L}\epsilon = -0.0410 + 8.103\times0.005 = -0.000485$, thus using Theorem \ref{th:guarantee}, we can guarantee the obtained $\delta$-ISS-CLF $V_{\theta,b}$ is valid and the closed-loop system is assured to be incrementally input-to-state stable under the influence of the controller $g_{\Bar{\theta},\Bar{b}}$.

The successful runs of the algorithm have an average convergence time of 2.5 hours.

\begin{figure}[h]
    \centering
    \includegraphics[width=0.6\linewidth]{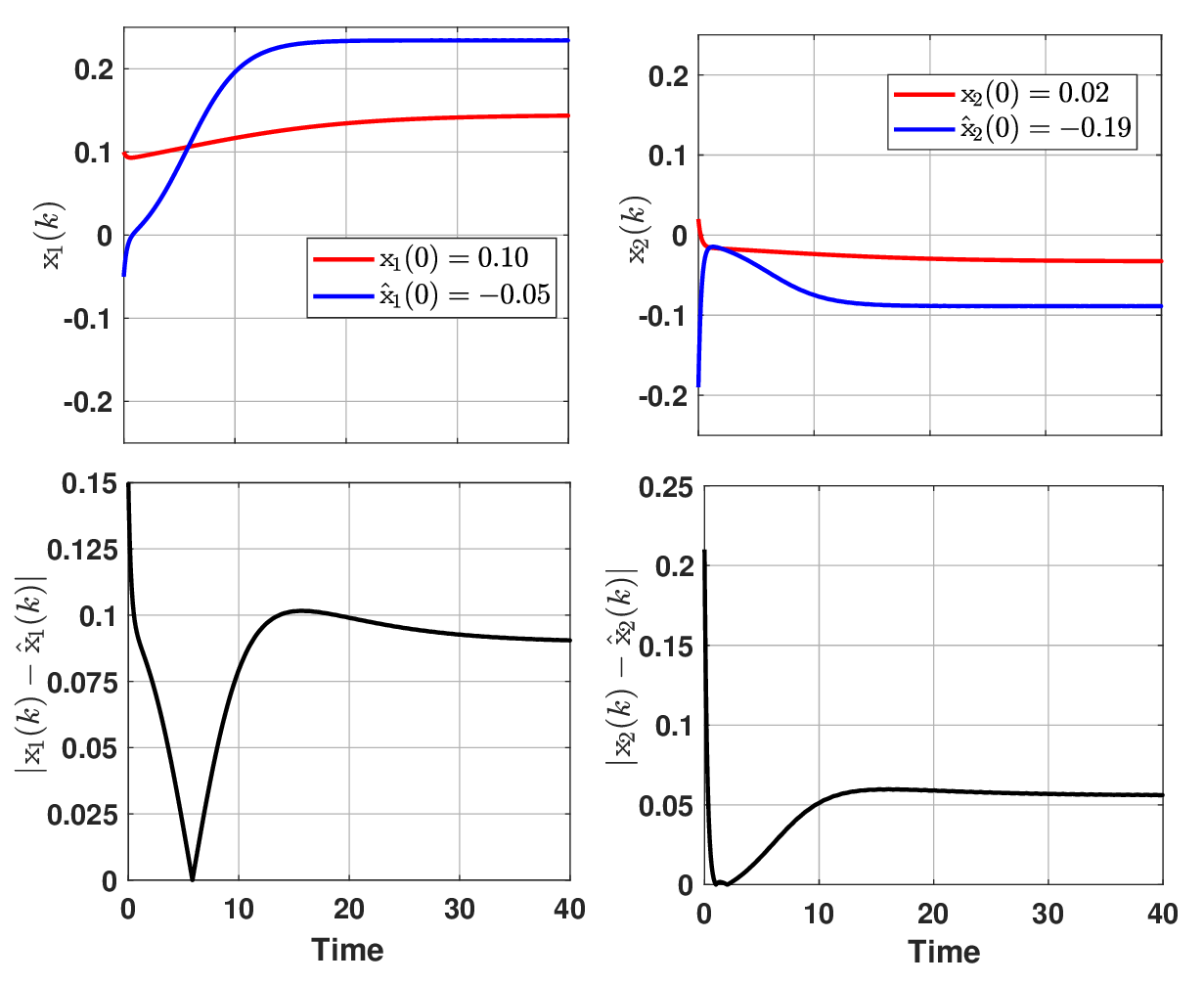}
    \caption{Top: Mass flow (left) and pressure rise (right) of the Jet-engine model, where the blue curve is influenced under input $\w(t) = -0.01 \in \W$, and the red curve is influenced under input $\w(t) = -0.1 \in \W$ for all $t \geq 0$. Bottom: The difference in mass flow (left) and pressure rise (right) subjected to different initial conditions and input flows through the throttle.}
    \label{fig:sim3}
\end{figure}

One can see from Figure \ref{fig:sim3} that under the different input conditions, the trajectories corresponding to various states starting from different initial conditions maintain the same distance or converge to a particular trajectory after some time instances under the influence of the controller.

\begin{figure*}[t]
    \centering
    \begin{subfigure}{0.32\textwidth}
        \centering
        \includegraphics[width=\textwidth]{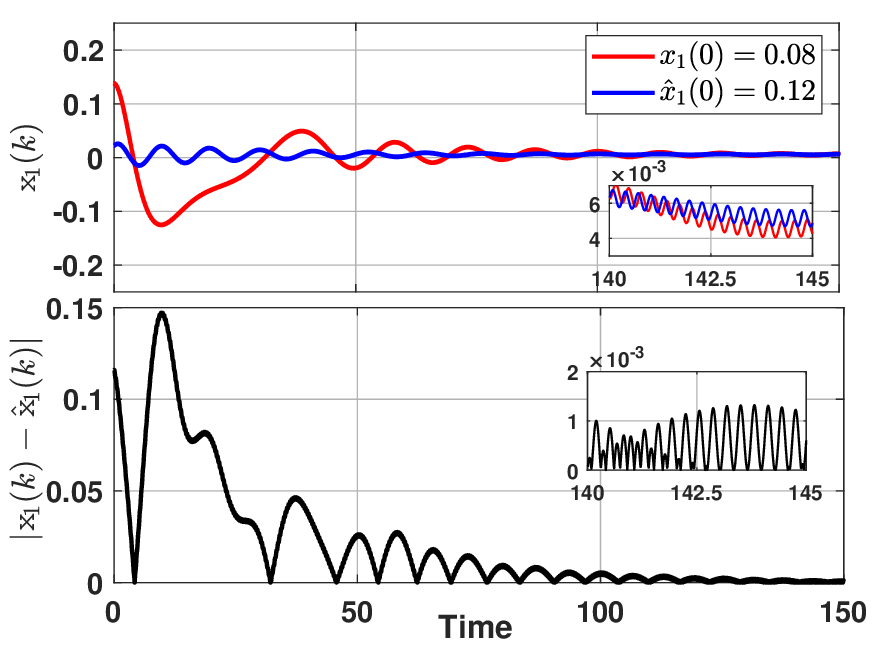}
        \caption{}
    \end{subfigure}
    \hfill
    \begin{subfigure}{0.32\textwidth}
        \centering
        \includegraphics[width=\textwidth]{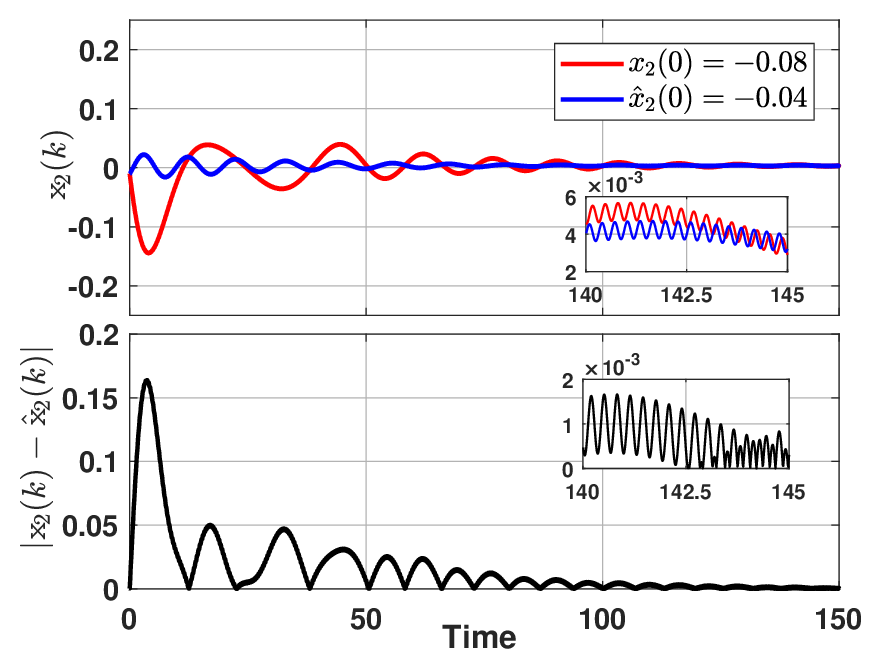}
        \caption{}
    \end{subfigure}
    \hfill
    \begin{subfigure}{0.32\textwidth}
        \centering
        \includegraphics[width=\textwidth]{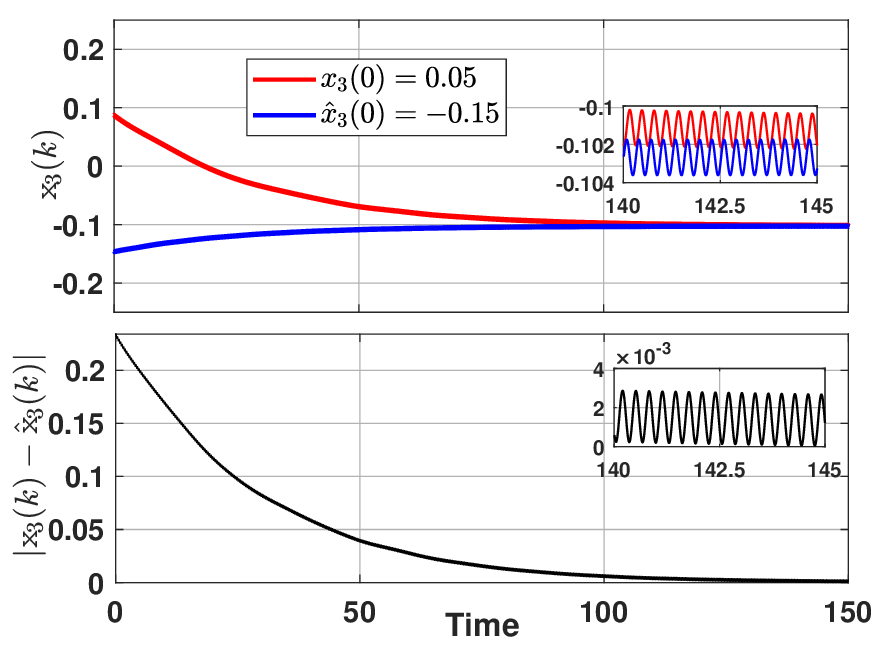}
        \caption{}
    \end{subfigure}
    \vspace{-0.2cm}
    \caption{Top: Trajectories ((a) $\omega_1$, (b) $\omega_2$, (c) $\omega_3$) starting from different initial conditions under different input signals (for all $k\in\N_0$, $\w(t)= \cos(2t)\in \W$ for blue curve and $\hat\w(t) = \sin(2t) \in \W$ for red curve), Bottom: The difference between the trajectories corresponding to different states.}
    \label{fig:sim6}
\end{figure*}

\subsection{Rotating Spacecraft Model}
We consider another example of a rotating rigid spacecraft model \cite{khalil2002nonlinear}, whose discrete-time dynamics is governed by the following set of equations.
\begin{align}
    \dot{\mathsf{x}}_1(t) &=  \frac{J_2 - J_3}{J_1} \mathsf{x}_2(t) \mathsf{x}_3(t) + \frac{1}{J_1} \mathsf{u}_1(t),\notag \\
    \dot{\mathsf{x}}_2(t) &= \frac{J_3 - J_1}{J_2} \mathsf{x}_1(t) \mathsf{x}_3(t) + \frac{1}{J_2} \mathsf{u}_2(t),\notag \\
    \dot{\mathsf{x}}_3(t) &= \frac{J_1 - J_2}{J_3} \mathsf{x}_1(t) \mathsf{x}_2(t) + \frac{1}{J_3} \mathsf{u}_3(t),
\end{align}
where $\mathsf{x} = [\mathsf{x}_1, \mathsf{x}_2, \mathsf{x}_3]^\top$ denotes angular velocities $\omega_1, \omega_2, \omega_3$ along the principal axes respectively,  $\mathsf{u} = [\mathsf{u}_1, \mathsf{u}_2, \mathsf{u}_3]^\top$ represents the torque input, and with $J_1 = 200, J_2 = 200, J_3 = 100$ denote the principal moments of inertia. We consider the state space of the system to be $\X = [-0.25,0.25]\times[-0.25,0.25]\times [-0.25,0.25]$. Moreover, we consider the input set to be bounded within $\W = [-10,10]$. Note that the internal input has 3 dimensions, while the external input has a single dimension,  that is, $\U \subset \R^3$, but $\W \subset \R$. In addition, we consider the model to be unknown. However, we estimate the Lipschitz constants $\mathcal{L}_x = 0.160, \mathcal{L}_u = 0.015$. 

The goal is to synthesize a controller to enforce the system to be $\delta$-ISS. To do this, we first fix the training hyper-parameters as $\epsilon = 0.0125, \mathcal{L}_L = 1, \mathcal{L}_{dL} = 1, \mathcal{L}_C = 20, k_1 = 0.00001, k_2 = 0.02, k_w = 0.01, \kappa = 0.0001, \mu_h = 0.01$. So, the Lipschitz constant according to Theorem \ref{th:constr} is $1.4142$.
We fix the structure of $V_{\theta,b}$ as $l_f = 2, h_f^1 = 60$ and $g_{\Bar{\theta},\Bar{b}}$ as $l_c = 1, h_c^1 = 40$. The activation functions for $\delta$-ISS-CLF and the controller are Softplus and ReLU, respectively.
The training algorithm converges to obtain $\delta$-ISS-CLF $V_{\theta,b}$ along with $\eta = -0.0180$. Hence, $\eta+\mathcal{L}\epsilon = -0.0182 + 1.4542\times0.0125 = -0.00002$, therefore, using Theorem \ref{th:guarantee}, we can guarantee that the obtained $\delta$-ISS-CLF $V_{\theta,b}$ is valid and that the closed-loop system is assured to be incrementally input-to-state stable under the influence of the controller $g_{\Bar{\theta},\Bar{b}}$.

The successful runs of the algorithm have an average convergence time of 5 hours.

One can see from Figure \ref{fig:sim6}(a), \ref{fig:sim6}(b), and \ref{fig:sim6}(c), under the different input conditions, the trajectories corresponding to various states starting from different initial conditions maintain the same distance after some time instances under the influence of the controller.

\subsection{Discussion and Comparisons} \label{subsec:Discuss}
The proposed approach to synthesizing controllers, ensuring incremental stability for closed-loop continuous-time systems, provides several benefits. The data-driven neural controller synthesis utilizes the data from the state-space of the system. Existing approaches for designing controllers that ensure incremental input-to-state stability, such as those in \cite{zamani2013backstepping,jagtap2017backstepping}, require full knowledge of the system dynamics and assume a specific structure, typically a strict-feedback form. A recent work \cite{sundarsingh2024backstepping} proposed estimating the strict feedback nature using the Gaussian process and designing a backstepping-like control to make the system incrementally stable. Compared to these methods, our approach can handle any general nonaffine nonlinear system, as seen in the Case Study \ref{Case study 1} of a general nonlinear system. Although recent advancement \cite{zaker2024controller} uses data collected from system trajectories to design a controller for an input-affine nonlinear system with polynomial dynamics, our method extends to arbitrary nonlinear dynamics, as seen in the first case study, where neither input-affine structures nor polynomial forms are present.

The final case study demonstrates that the system possesses three control signals, but can be effectively controlled using a reduced number of external inputs. Thereby, the proposed methodology is able to make the closed-loop system controllable with a reduced number of inputs, simplifying the control implementation. This will be particularly beneficial in improving the scalability of abstraction-based controller synthesis approaches \cite{sadek2023compositional,symbolic1,jagtap2017quest}. As discussed in \cite{zamani2017towards}, the computational complexity of constructing an input-based abstraction of an incrementally stable system depends on the input dimension of the system. As we are able to make the system incrementally stable with a reduced number of input dimensions, the complexity of the abstraction is significantly reduced. This facilitates the efficient synthesis of additional formally verified controllers to meet other specifications for the closed-loop system.


\section{Conclusion and Future Work}
This study shows how to synthesize a formally validated neural controller that ensures incremental input-to-state stability for the closed-loop system. The existence of an $\delta$-ISS-CLF guarantees $\delta$-ISS of the closed-loop system. The constraints of $\delta$-ISS control Lyapunov function are formulated into an ROP, then collecting data from sample space and mapping the problem into SOP. A validity condition is proposed that guarantees a successful solution of the SOP results in satisfying the ROP as well, thereby resulting in a valid $\delta$-ISS control Lyapunov function. The training framework is then proposed, utilizing the validity condition, to synthesize the provably accurate $\delta$-ISS-CLF under the controller's action and formally ensure its validity by changing appropriate loss functions.

The work can be extended into constructing $\delta$ISS Control Lyapunov functions under the action of the controller for interconnected systems as well, which is currently being investigated. We will try to synthesize $\delta$-ISS Lyapunov functions for each subsystem and claim how they can be used to formally guarantee the incremental stability of the large system. In addition, future direction of this work can be in the purview of stochastic systems, providing a formal guarantee of their incremental stability.

\bibliographystyle{plain}
\bibliography{sources}

\appendix

\section{Proof of Theorem \ref{th:admit}} \label{appendix:admit}
The proof is inspired by the proof of Theorem 2.6 of \cite{zamani2011lyapunov}. Let $\mathsf{x}_{x,\mathsf{w}}(t)$ denotes the state of the closed-loop system at time $t$ under input signal $\mathsf{w}$ from the initial condition $x=\mathsf{x}_{x,\mathsf{w}}(0)$. Note that, since the system is forward invariant under the action of the controller $g_{\bar{\theta}, \bar{b}}$, for all $t \geq 0$ we have $\mathsf{x}_{x,\mathsf{w}}(t) \in \X$. Consider the closed-loop system admits a $\delta$-ISS-CLF under a controller. Now, by using property (i) of Definition \ref{def:ISS-Lf}, we get
\begin{align}\label{eq:app_admit_1}
    |\mathsf{x}_{x, \mathsf{w}}(t) - \mathsf{x}_{\hat{x}, \hat{\mathsf{w}}}(t)| \leq \alpha_1^{-1}(V(\mathsf{x}_{x, \mathsf{w}}(t), \mathsf{x}_{\hat{x}, \hat{\mathsf{w}}}(t))),
\end{align}
for any $t \in \R_0^+$. Now using condition (ii) of Definition \ref{def:ISS-Lf} and the comparison lemma \cite[Lemma 3.4]{khalil2002nonlinear}, we get 
\begin{align}\label{eq:app_admit_2}
    V(\mathsf{x}_{x, \mathsf{w}}(t), \mathsf{x}_{\hat{x}, \hat{\mathsf{w}}}(t)) \leq e^{-\kappa t}V(x, \hat{x}) + e^{-\kappa t}* \sigma(|\mathsf{w}(t) - \hat{\mathsf{w}}(t)|),
\end{align}
for any $t \in \R_0^+$, and $*$ denotes the convolution integral. Now combining \eqref{eq:app_admit_1} and \eqref{eq:app_admit_2}, we get
\begin{align*}
    & |\mathsf{x}_{x, \mathsf{w}}(t) - \mathsf{x}_{\hat{x}, \hat{\mathsf{w}}}(t)| \leq \alpha_1^{-1} \bigg(e^{-\kappa t}V(x, \hat{x}) + e^{-\kappa t}* \sigma(|\mathsf{w}(t) - \hat{\mathsf{w}}(t)|)\bigg) \\
    & \leq \alpha_1^{-1} \bigg(e^{-\kappa t}V(x, \hat{x}) + \frac{1 - e^{-\kappa t}}{\kappa} \sigma(\lVert \mathsf{w} - \hat{\mathsf{w}}\rVert)\bigg) \\
    & \leq \alpha_1^{-1} \bigg(e^{-\kappa t}V(x, \hat{x}) + \frac{1}{\kappa} \sigma(\lVert \mathsf{w} - \hat{\mathsf{w}}\rVert)\bigg) \\
    & \leq \alpha_1^{-1} \bigg(e^{-\kappa t}\alpha_2(|x - \hat{x}|) + \frac{1}{\kappa} \sigma(\lVert \mathsf{w} - \hat{\mathsf{w}}\rVert)\bigg) := \Psi(\rho, \Upsilon),
\end{align*}
where $\Psi(\rho, \Upsilon) := \alpha_1^{-1}(\rho + \Upsilon), \rho := e^{-\kappa t}\alpha_2(|x - \hat{x}|), \Upsilon:= \frac{1}{\kappa} \sigma(\lVert \mathsf{w} - \hat{\mathsf{w}}\rVert)$. Notice that $\Psi$ is nondecreasing in both variables, we get:
\begin{align*}
    |\mathsf{x}_{x, \mathsf{w}}(t) - \mathsf{x}_{\hat{x}, \hat{\mathsf{w}}}(t)| \leq \xi(e^{-\kappa t}\alpha_2(|x - \hat{x}|)) + \xi(\frac{1}{\kappa} \sigma(\lVert \mathsf{w} - \hat{\mathsf{w}}\rVert)),
\end{align*}
where $\xi(r) = \Psi(r,r) = \alpha_1^{-1}(2r)$ and $\xi:\R_0^+ \rightarrow \R_0^+$ is a class $\mathcal{K}_{\infty}$ function. Then finally,
\begin{align*}
    |\mathsf{x}_{x, \mathsf{w}}(t) - \mathsf{x}_{\hat{x}, \hat{\mathsf{w}}}(t)| &\leq \alpha_1^{-1}(2e^{-\kappa t}\alpha_2(|x - \hat{x}|)) + \alpha_1^{-1}(\frac{2}{\kappa} \sigma(\lVert \mathsf{w} - \hat{\mathsf{w}}\rVert)).
\end{align*}
Hence, defining 
\begin{align}
    &\beta(|x - \hat{x}|,t):= \alpha_1^{-1}(2e^{-\kappa t}\alpha_2(|x - \hat{x}|)), \\
    &\gamma(\lVert \mathsf{w} - \hat{\mathsf{w}}\rVert):= \alpha_1^{-1}(\frac{2}{\kappa} \sigma(\lVert \mathsf{w} - \hat{\mathsf{w}}\rVert)),
\end{align}
we can conclude the closed-loop system will be $\delta$-ISS under the action of the controller within the state space.

\section{Proof of Lemma \ref{lem:cfi_guarantee}} \label{appendix:cfi}
Consider a function $h: \X \rightarrow \R$ exists such that the conditions of \eqref{eq:leq_BC} holds,  \textit{i.e.}, $h(x) = 0, \forall x \in \partial \X$ and $h(x) > 0, \forall x \in int(\X)$.
    
Now, we assume that there exists a controller $g:\X \rightarrow \U$ such that condition \eqref{eq:diff_BC} is satisfied. Then, one can infer at the boundary of $\X$, the derivative of $h$ is always positive, enforcing the trajectories of the system under the controller $g$ to move towards inside the state space. Hence, the state space $\X$ becomes control forward invariant. This completes the proof.

\section{Proof of Lemma \ref{lem:Lipschitz_lie}} \label{appendix:Lipschitz_lie}
{From the definition of $\mathsf{L}_{f,g}V(x, \hat{x})$ and using Assumption \ref{assum:Lipschitz_net}, \ref{assum:bound}, we can have, for any $x_1,x_2, \hat{x}_1, \hat{x}_2 \in \X$ and $w_1, w_2, \hat{w}_1, \hat{w}_2 \in \W$: \\
$|\mathsf{L}_{f,g}V(x_1, x_2) - \mathsf{L}_{f,g}V(\hat{x}_1, \hat{x}_2)| \leq 2|\frac{\partial V_{\theta,b}}{\partial x_1}f(x_1, g_{\theta,b}(x_1,w_1)) - \frac{\partial V_{\theta,b}}{\partial \hat{x}_1}f(\hat{x}_1, g_{\theta,b}(\hat{x}_1,\hat{w}_1))| \leq 2\big( \mathcal{M}_f \mathcal{L}_{dL}|x_1 - \hat{x}_1| + \mathcal{M}_L|f(x_1, g_{\theta,b}(x_1,w_1)) - f(\hat{x}_1, g_{\theta,b}(\hat{x}_1,\hat{w}_1))|\big)$. \\
Now, using Assumption \ref{assum:Lipschitz_fun} and \ref{assum:Lipschitz_net}, we have $|f(x_1, g_{\theta,b}(x_1,w_1)) - f(\hat{x}_1, g_{\theta,b}(\hat{x}_1,\hat{w}_1))| \leq |f(x_1, g_{\theta,b}(x_1,w_1)) - f(x_1, g_{\theta,b}(x_1,\hat{w}_1)) + f(x_1, g_{\theta,b}(x_1,\hat{w}_1)) - f(\hat{x}_1, g_{\theta,b}(\hat{x}_1,\hat{w}_1))| \leq (\mathcal{L}_x + \mathcal{L}_u\mathcal{L}_c)|x_1 - \hat{x}_1| + \mathcal{L}_u\mathcal{L}_c|w_1 - \hat{w}_1|$. \\
Therefore, combining them, we can conclude that $\mathsf{L}_{f,g}V(x, \hat{x})$ is Lipschitz continuous with Lipschitz constants $\mathscr{L}_{Vx}$ in $x$ and $\mathscr{L}_{Vu}$ in $w$ where $\mathscr{L}_{Vx} := 2\mathcal{M}_L(\mathcal{L}_x + \mathcal{L}_u\mathcal{L}_c) + 2\mathcal{M}_f\mathcal{L}_{dL}$ and $\mathscr{L}_{Vu}:=2\mathcal{M}_L\mathcal{L}_u\mathcal{L}_c$, thereby completing the proof.}

\section{Proof of Theorem \ref{th:delta_lie}} \label{appendix:delta}
{
With a slight abuse of notation, we denote $x_\mathsf{w}^t:=\mathsf{x}_{x, \mathsf{w}}(t)$ as the solution of the system \eqref{eq:system} at time $t \geq 0$ under the input sequence $\mathsf{w}$ starting from the initial condition $x$. Therefore, we can write,
\begin{align}\label{eq:Lyap_traj}
    V(x_\mathsf{w}^\tau,\hat{x}_{\hat{\mathsf{w}}}^\tau) = V(x, \hat{x}) + \int_0^\tau \mathsf{L}_{f,g}V(x_\mathsf{w}^t, \hat{x}_{\hat{\mathsf{w}}}^t)dt.
\end{align}
Now, considering the approximation $\widehat{\mathsf{L}}_{f,g}V(x, \hat{x})$ as in \eqref{eq:lie_approx}, combining it with \eqref{eq:Lyap_traj} and subtracting $\mathsf{L}_{f,g}V(x, \hat{x})$ from it, one can have
\begin{align*}
    & |\widehat{\mathsf{L}}_{f,g}V(x, \hat{x}) - \mathsf{L}_{f,g}V(x, \hat{x})| \\
    &\leq \frac{1}{\tau}\int_0^\tau |\mathsf{L}_{f,g}V(x_\mathsf{w}^t, \hat{x}_{\hat{\mathsf{w}}}^t) - \mathsf{L}_{f,g}V(x, \hat{x})|dt \\
    &\leq \frac{1}{\tau} \int_0^\tau \mathscr{L}_{Vx} (|x_\mathsf{w}^t - x| + |\hat{x}_{\hat{\mathsf{w}}}^t - \hat{x}|)dt \quad (\text{using Lemma \ref{lem:Lipschitz_lie}})\\
    &\leq \frac{\mathscr{L}_{Vx}}{\tau} \int_0^\tau (|x_\mathsf{w}^t - x| + |\hat{x}_{\hat{\mathsf{w}}}^t - \hat{x}|)dt.    
\end{align*}
Now we aim at finding the upper bound of $|x_\mathsf{w}^t - x|$. Under the continuity property of the solution of the system, we have $x_\mathsf{w}^t = x + \int_0^t f(x, g(x, \mathsf{w}(s)))ds$. Now, using Assumption \ref{assum:bound}, we can claim easily $|x_\mathsf{w}^t - x| \leq \int_0^t\mathcal{M}_fds = \mathcal{M}_ft$. Similarly, we can also say, $|\hat{x}_{\hat{\mathsf{w}}}^t - \hat{x}| \leq \mathcal{M}_ft$. Therefore, replacing them back to the previous equation, we have
\begin{align*}
    & |\widehat{\mathsf{L}}_{f,g}V(x, \hat{x}) - \mathsf{L}_{f,g}V(x, \hat{x})| \leq \frac{\mathscr{L}_{Vx}}{\tau} \int_0^\tau \mathcal{M}_ftdt = \tau \mathscr{L}_{Vx} \mathcal{M}_f,
\end{align*}
which completes the proof.}

\section{Proof of Theorem \ref{th:constr}} \label{appendix:constr}
{First, we prove that condition \eqref{eq:geq_SOP} with $\hat{\eta} + \mathcal{L}\varepsilon \le 0$ implies the satisfaction of condition \eqref{eq:geq}. Consider $x, \hat{x} \in \X$, then $V_{\theta,b}(x, \hat{x}) - k_1|x - \hat{x}|^{\gamma_1} \leq V_{\theta,b}(x, \hat{x}) - V_{\theta}(x_q, x_r) + V_{\theta}(x_q, x_r) - k_1|x - \hat{x}|^{\gamma_1} + k_1|x_q - x_r|^{\gamma_1} - k_1|x_q - x_r|^{\gamma_1} \leq \sqrt{2}\mathcal{L}_L\varepsilon + 2\mathsf{L}_1\varepsilon + \hat{\eta} \leq \hat{\eta} + \mathcal{L}\varepsilon \leq 0$, showcasing the satisfaction of condition \eqref{eq:geq}. Similarly, one can follow these steps for the proof of satisfaction of other conditions, which implies that if we consider such $\hat{\eta}$ that satisfies $\hat{\eta} + \mathcal{L}\varepsilon \le 0$, the lemma \ref{lem:ROP} has been satisfied, thereby establishing $V_{\theta,b}$ to be a valid $\delta$-ISS-CLF for the system \eqref{eq:system} under the controller $g_{\bar{\theta}, \bar{b}}$.}

\section{Proof of Theorem \ref{th:derivative}} \label{appendix:derivative}
Let the dimension of the input $x$ be $r \times 1$, the dimension of the zeroth weight $\theta_0$ be $p \times r$, the first weight $\theta_1$ be $q \times p$ and so on, the final weight $\theta_{\No}$ is of dimension $1 \times s$ and the output is scalar. 

Let $\phi_1, \phi_2, \cdots, \phi_{\No}$ be the activation functions of the layers of the neural network.
So, the NN is given by:
\begin{align*}
    y &= \theta_{\No} \phi_{\No}(\theta_{\No-1} \phi_{\No-1}(\cdots \phi_1 (\theta_0x+b_0)\cdots)+b_{\No-1}) \\
    \frac{\partial y}{\partial x} &= \theta_{\No}\text{diag}(\phi_{\No}')\theta_{\No-1}\text{diag}(\phi_{\No-1}')\ldots\theta_1 \text{diag}(\phi_1')\theta_0,
\end{align*}
where $\phi_i':= \phi_i'(\theta_{i-1}\phi_{i-1}(\ldots \phi_1(\theta_0x + b_0)\ldots)+b_{i-1})$ for all $i \in \{1,\ldots,\No\}$. Now following the Assumption \ref{assum:lipschitz_act_fun}, we can infer $\phi_i' \leq \Lo_i$. So, we replace the derivatives by corresponding Lipschitz bounds except $\phi_{\No}'$. Hence, the derivative will look like:
\begin{align*}
    \frac{\partial y}{\partial x} &\leq \theta_{\No}\text{diag}(\phi_{\No}')\theta_{\No-1}\Lo_{\No-1}\mathcal{I}_{\No-1}\ldots\theta_1 \Lo_1\mathcal{I}_{1}\theta_0 \\
    &\leq (\Lo_1 \ldots \Lo_{\No-1})\theta_{\No}\text{diag}(\phi_{\No}')\theta_{\No-1} \ldots \theta_0,
\end{align*}
where $\mathcal{I}_i, i \in \{1, \ldots, (\No-1)\}$ is identity matrices with appropriate dimensions. Let us consider $\Lo_1 \ldots \Lo_{\No-1} = \Lo$. Also, as the dimension of $\frac{\partial y}{\partial x}$ is $1 \times r$, its transpose will have the dimension of $r \times 1$.
\begin{align*}
    (\frac{\partial y}{\partial x})^\top &\leq \Lo(\theta_{\No}\text{diag}(\phi_{\No}')\theta_{\No-1} \ldots \theta_0)^\top.
\end{align*}
Now, since $\theta_{\No}$ and $\phi_{\No}'$ are vectors of the same dimension, we can say $\theta_{\No}\text{diag}(\phi_{\No}') = (\phi_{\No}')^\top \text{diag}(\theta_{\No})$. Hence,
\begin{align*}
    (\frac{\partial y}{\partial x})^\top &\leq \underbrace{\Lo\theta_0^\top\theta_1^\top\ldots\theta_{\No-1}^\top\text{diag}(\theta_{\No})}_{\hat{\theta}_{\No}}\phi_{\No}'.
\end{align*}
So one can upper bound the derivative of the original network as a neural network. Hence, the Lipschitz continuity of the derivative network can be enforced by $M_{\mathcal{L}_{dL}}(\hat{\theta}, \hat{\Lambda}) \geq 0$ where $\hat{\theta}$ is as mentioned in the theorem with $\mathcal{L}_{dL}$ being the Lipschitz constant. This completes the proof.

\end{document}